# Evaluation of adaptive treatment strategies in an observational study where time-varying covariates are not monitored systematically

Noémi Kreif[1], Oleg Sofrygin[2], Julie Schmittdiel[3], Alyce Adams[4], Richard Grant[5], Zheng Zhu[6], Mark van der Laan[7], and Romain Neugebauer [*][8]

[1]Centre for Health Economics, University of York, York, UK

[2,3,4,5,6,8]Kaiser Permanente Northern California, Division of Research, Oakland, CA, USA

[2,7]Division of Biostatistics, School of Public Health, University of California, Berkeley, CA, USA

June 28, 2018

*corresponding author details: romain.s.neugebauer@kp.org. This is an independent report partially funded by the Medical Research Council (Early Career Fellowship in the Economics of Health, Dr Noemi Kreif MR/L012332/1). This study was also partially supported through a Patient-Centered Outcomes Research Institute (PCORI) Award (ME-1403-12506). All statements in this report, including its findings and conclusions, are solely those of the authors and do not necessarily represent the views of the Patient-Centered Outcomes Research Institute (PCORI), its Board of Governors or Methodology Committee. The authors thank the following investigators from the HMO research

## Abstract

In studies based on electronic health records (EHR), the frequency of covariate monitoring can vary by covariate type, across patients, and over time. This can lead to major challenges: first, the difference in monitoring protocols may invalidate the extrapolation of study results obtained in one population to the other, and second, monitoring can act as a time-varying confounder of the causal effect of a time-varying treatment on the outcomes of interest. This paper demonstrates how to account for non-systematic covariate monitoring when evaluating dynamic treatment interventions, and how to evaluate joint dynamic treatment-censoring and static monitoring interventions, in a real world, EHR-based, comparative effectiveness research (CER) study of patients with type II diabetes mellitus. First, we show that the effects of dynamic treatment-censoring regimes can be identified by including indicators of monitoring events in the adjustment set. Second, we demonstrate the poor performance of the standard inverse probability weighting (IPW) estimator of the effects of joint treatment-censoring-monitoring interventions, due to a large decrease in data support resulting in a large increase in standard errors and concerns over finite-sample bias from near-violations of the positivity assumption for the monitoring process. Finally, we detail an alternate IPW estimator of the effects of these interventions using the No Direct Effect assumption. We demonstrate that this estimator can result in improved efficiency but at the cost of increased bias concerns over structural near-violations of the positivity assumption for the treatment process. To conclude, this paper develops and illustrates new tools that researchers can exploit to appropriately account for non-systematic covariate monitoring in CER, and to ask new causal questions about the joint effects of treatment and monitoring interventions.

# 1    Introduction

## 1.1    Dynamic treatment regimes and non-systematic covariate monitoring in longitudinal studies

In the management of chronic conditions such as diabetes, adaptive treatment strategies (a.k.a. dynamic treatment regimes) might often lead to better outcomes than non-adaptive treatment strategies (a.k.a. static treatment regimes).[1] In addition, adaptive treatment strategies better reflect real-world decision making because they allow for the update of treatment decisions over time as a function (hereinafter, referred to as a decision rule) of the changing circumstances (e.g. prognosis) of the individual patient. Dynamic treatment regimes can be evaluated using randomised experiments (SMART trials, see e.g. (2)) and increasingly in comparative effectiveness research, using observational studies.[3,4,5]

Regardless of the study design, evaluating adaptive treatment strategies requires data collection on covariates that enter the decision rules of interest. For example, for the treatment of HIV, adaptive treatment strategies such as "start treatment when CD4 count first drops below 350" require that patient's CD4 count be measured over time. The pattern and frequency of covariate measurements vary across study designs. In randomised trials, covariates are typically measured at regular intervals but the lengths of intervals between monitoring events can also vary over time and by covariate type within the same experiment.[6] In widely employed observational study designs (e.g. retrospective cohort studies based on electronic health records[7]), monitoring is typically a joint decision of the patient and the health care providers, and is thus not expected to be synchronized between patients as would be the case in observational studies (e.g., some prospective cohort studies) in which investigators exert more direct control on the data collection protocols.

Thus, the frequency of covariate measurements can greatly vary by covariate type, across patients, and over time in observational data derived from EHR-based cohort studies. For example, in two studies of antiretroviral treatments, (8) collect covariate information every 7 days, while (9) use a 180-day intervals of measurement. This monitoring variability can lead to challenges but also new opportunities for research that is based on the evaluation of adaptive treatment strategies for the following three reasons.

First, when aiming to establish the optimal dynamic treatment regime, optimality will likely depend on the frequency of monitoring.[9] For example, in a diabetes population where hemoglobin A1c (A1c) is monitored every year, the optimal A1c threshold for treatment intensification (e.g., addition of a drug to protect against microvascular complications) might differ from that in a population where A1C is monitored every 3 months. One might expect that, in the first population, a more aggressive treatment strategy (i.e., treatment intensification at a lower A1c level) is necessary to avoid the adverse health outcomes from delays in treatment intensification that would be caused by infrequent clinical monitoring if patients were treated according to the optimal A1c threshold for treatment intensification in the second population.[9] Hence, the difference in monitoring protocols between two populations may invalidate the extrapolation of study results obtained in one population to the other.

Second, monitoring in itself is a health intervention with non-negligible costs to healthcare systems that can also burden patients financially or otherwise (e.g., monitoring can cause discomfort or adverse events in case of invasive procedures such as biopsies). Hence establishing the optimal monitoring intervention may be of interest in its own right. As illustrated in the previous paragraph, evaluating joint interventions that intervene on both the monitoring regime (for example, "monitor every 6 months") and on the treatment regime is also of particular interest to enhance the generalizability of study findings when evaluating adaptive treatment strategies.

Third, monitoring can act as a time-varying confounder of the causal effect of a time-varying treatment on the outcomes of interest. Patients with frequent monitoring events may be very different from patients who rarely interact with the health care system, with respect to (possibly unobserved) characteristics such as disease severity or health related behaviour, and these characteristics may also impact patient outcomes. Hence, in observational studies and in particular those based on retrospective cohort studies, accurate estimation of the causal effect of dynamic treatment regimes may often require adjustment for the occurrence of monitoring events.

While statistical methods for estimating the causal effects of dynamic treatment regimes are well developed,[10,3,11,12,9,13,14] there is little practical guidance on how to handle the challenges of non-systematic monitoring and how to exploit monitoring variability to evaluate its impact on health outcomes and improve the generalizability of study results. (9) develop estimators for the joint optimal treatment-monitoring strategies. One particular issue highlighted by (9) is the requirement for a positivity assumption about the monitoring process that generated the study data: more specifically, conditional on a subject's past observed variables, it is required that there is a positive probability for that subject to follow various monitoring regimes of differing monitoring intervals (e.g., monitoring every 3 months versus every 3 years). This requirement, in real world settings, may often be implausible. The estimator they propose weakens the positivity assumption requirement for monitoring by introducing the "no direct effect" (NDE) assumption that requires that the monitoring decision can only impact the outcome through subsequent treatment decisions, for example excluding the possibility of monitoring affecting diet and physical activity which, in turn, would impact health outcomes. (15) build on these results within the nonparametric structural equation modeling framework by deriving new identifiability results that facilitate the construction of estimators of effects defined by general joint treatment and monitoring interventions, using the same NDE assumption.

Beyond (6), who apply some of the methods proposed by (9) to estimate the joint effect of CD4 monitoring and antiretroviral therapy initiation strategies in HIV infected adults, there are, to our knowledge, no published studies that explore the challenges posed by non-systematic monitoring in a real world case study. As observational studies make increasing use of routinely collected longitudinal data sources which were not originally intended for research purposes (e.g., electronic health records data), the development of methods and practical guidelines for handling non-systematic monitoring in the evaluation of longitudinal treatment regimes is necessary. This paper demonstrates how to tackle the challenges and opportunities just described in a real world case study based on electronic health records (EHR) data.

## 1.2 A case study to illustrate and evaluate the practical performance of IPW estimation to control for non-systematic covariate monitoring: the TI study

We build on a retrospective cohort study designed to evaluate progressively more aggressive glucose-lowering strategies on clinical outcomes of adults with type II diabetes mellitus.[16,17] The study is based on a large, longitudinal, EHR data set of 51,179 patients from the Health Maintenance Organization (HMO) Research Network.[18] The dynamic treatment interven-tions under consideration require that a patient initiates treatment intensification (TI) at the first time her A1c level reaches or drifts above $x\%$ and that the patient remains on the intensified therapy thereafter, with $x = 7$, 7.5, 8, or 8.5. To appropriately account for time-dependent confounding and informative censoring, dynamic marginal structural modelling was employed, using inverse probability weighting (IPW) and targeted minimum loss based estimation.[19, 20, 21] Findings from this observational study demonstrated the benefit of tight glycemic control with respect to the development or progression of albuminuria.

We build on the various methodological approaches applied to date in this study by now addressing the problem of non-systematic monitoring when estimating the comparative effectiveness of the four treatment intensification strategies described above. We first revisit how prior IPW estimation were implemented to discuss an assumption about monitoring events that has, to our knowledge, not been highlighted in the literature despite its common but implicit adoption in prior work on the evaluation of dynamic treatment regimes. Here, we relax this assumption and describe practical implications for constructing consistent IPW estimators of the effects of dynamic treatment regimes. Second, we describe the standard IPW estimation of the effects of joint treatment-censoring-monitoring interventions (for example, monitor A1c every 6 months, and intensify treatment when A1c first exceeds 7.5%). Finally, we detail an alternate approach that was proposed to handle the practical limitations expected to arise when implementing this standard IPW estimator. More specifically, using identification results from (15), we describe an alternate IPW estimator of the effects of joint monitoring-treatment interventions using the no direct effect assumption. In each section, we illustrate and evaluate the practical performance of the analysis described using data from the aforementioned diabetes study.

For each analysis, we describe the causal estimands, their identifiability assumptions, a distinct IPW estimator, and its implementation for evaluating the causal estimands. We present results in terms of both estimated counterfactual survival curves and estimated counterfactual cumulative risk differences. Throughout we use IPW estimators of the counterfactual hazards whose estimates are then mapped into survival probability estimates. All analyses were implemented using the `stremr` R package.[22]

## 1.3   Observed data notation

While EHR data are captured in continuous time (i.e. information such as medication dispensed or results from lab exams are stamped with event dates and times), it is common practice to first coarsen granular EHR data by mapping them into the organized structure described below using a discrete time scale chosen by the analyst. This data structure is motivated by the widely used framework developed in the literature for effect estimation with complex observational longitudinal data such as the evaluation of dynamic treatment effects.[23, 11, 9] All analyses in this report are based on an analytic dataset constructed with the `MSMstructure` SAS macro[24] to coarsen daily EHR data using the 90-day time unit. We now describe the random variables whose realizations on each of 51,179 patients make up the analytic data set.

For each patient in the cohort, measurements on treatment, covariate, censoring and outcome information are updated every 90 days starting at study entry and until the end of follow-up. Follow-up time (denoted by $t$) is thus expressed in 90-day units. By convention, the first 90 days of follow-up is denoted by $t = 0$. The end of follow-up is the earliest of the time to the failure event (i.e., albuminaria development or progression), or the time to a censoring event (i.e., death, disenrollment from the health plan, or administrative end of study). The latest possible follow-up time (expressed in 90-day units) is denoted by $K + 1 = 36$, corresponding to about 9 years. The binary outcome variable $Y(t)$ represents whether the failure event occurred at the previous time point $t - 1$, and $Y(0) = 0$ by convention. The vector $A(t) = (A_1(t), A_2(t))$ denotes the exposure status at time $t$ and is composed of two binary variables $A_1(t)$ and $A_2(t)$ that indicate whether the patient experienced an intensified therapy and a censoring event at time $t$, respectively.

The vector of covariates $Z(t)$ represents various patient attributes measured before $A(t)$ (e.g., comorbidity diagnoses or vital signs). The covariate $I(t)$ denotes the patient attribute measured before $A(t)$ that is involved in the definition of the dynamic treatment regimes of interest, i.e., the A1c level. This is also the covariate whose monitoring frequency plays a central role in the evaluation of the dynamic treatment regimes and for which the evaluation of monitoring interventions also becomes of interest. Note that $I(t)$ is not necessarily monitored at every $t > 0$ for all patients in this study. The binary variable $N(t)$ represents the monitoring decision at time $t$ and indicates whether a measurement for $I(t+1)$ will be taken. By convention, when A1c is not monitored at time $t$, $I(t)$ is defined as $I(j)$ where $j < t$ is the latest time point when A1c was monitored. By study design, in the first period A1c is always monitored. We jointly denote the outcome and covariates at time $t$ by the vector $L(t) \equiv (Y(t), Z(t), I(t))$ which, by definition, consists of measurements obtained before the exposure $A(t)$.

To sum-up, the observed data are realizations of $n$ copies $O_i$ of the following temporally ordered sequence of random variables:

$$O = (L(0), A(0), N(0), ...., L(K), A(K), N(K), L(K+1))$$

where, by convention, all variables become degenerate after a failure or censoring event occurs. Following typical practice, the analyses in this report assume that the random variables $O_i$ are independent and identically distributed. To simplify notations below, we use the overbar to denote the history of a variable. For example, $\bar{Z}(t)$ denotes the history of covariate $Z$ from baseline to time $t$, i.e., $\bar{Z}(t) = (Z(0), \dots, Z(t))$. By convention, $L(t)$, $A(t)$, and $N(t)$ are nil when $t < 0$.

## 1.4 Causal Model

We assume a nonparametric structural equation model (NPSEM)[25, 26] linking the observed data distribution to a vector of random disturbances and a fixed vector of functions, described in detail in (15). In this NPSEM, the observed level of the A1c, $I(t)$, is linked to a latent variable $I^0(t)$ that represents the patient's underlying A1c level at time $t$. More specifically, if a decision to measure $I^0(t)$ was made during follow-up (i.e., $N(t-1) = 1$), we then observe $I(t)$ to be $I^0(t)$, and otherwise the value of $I^0(t)$ is missing from the observed data $O$. Because A1c is always monitored in the first period, we have $I(0) = I^0(0)$ for all patients. This NPSEM also defines all counterfactual outcomes[27] whose distributions define the various causal estimands of interest in this report.

## 2 Evaluation of dynamic treatment regimes: a review to highlight special considerations for studies with non-systematic clinical monitoring

### 2.1 Causal estimands

We consider four distinct longitudinal treatment-censoring interventions which each sets the random variable $A(t) = (A_1(t), A_2(t))$ to a possible realisation $a(t) = (a_1(t), a_2(t))$ for each time period, $t = 0, ..., t_0$ where $t_0 < K$ is the time point chosen by the analyst when outcomes are compared between intervention groups. Specifically, we focus on dynamic treatment regimes where the values of $a_1(t)$ are not pre-defined (such as "always treat" or "never treat") but adapted to time-varying realizations of patient characteristics based on the following decision rules applied through failure occurrence: "intensify diabetes treatment the first time A1c reaches or drifts above x%, and keep patient on the intensified therapy thereafter" with x=7;7.5;8;8.5. In addition, the four interventions of interest also require that patients do not experience a censoring event, i.e., $\bar{a}_2(t_0) = 0$, to ensure that the outcome at $t_0$ is always observed.

Formally, each of the four interventions above requires to set the treatment value to $a_1(t) = 1$ (i.e., initiate an intensified treatment) for a patient who has not failed yet (i.e., $Y(t) = 0$) and who has not received intensified treatment before, $(\bar{A}_1(t-1) = 0)$ but for whom a new measurement of the A1C has been taken $(N(t-1) = 1)$ and its level has reached or exceeded a certain threshold $I(t) \geq x$. Otherwise, for a patient who has not yet failed, and who has also not received an intensified treatment before and for whom a new A1c measurement has been taken $(N(t-1) = 1)$ but its level $I(t)$ has not reached or exceeded the threshold, the intervention requires that this patient not initiate an intensified treatment (set $a_1$ to 0). For a patient who has not failed yet and who, either, did not undergo a new A1c test ($N(t-1) = 0$), or had already initiated an intensified treatment previously ($\bar{A}(t-1) = 1$), then the intervention requires that treatment be not changed, i.e $a_1(t) = a_1(t-1)$, corresponding to "no treatment initiation" if the patient was not on intensified treatment before, and to "continuation of the intensified treatment" if the patient had already initiated it. Each intervention also requires to set the censoring variable $a_2(t)$ to value 0 in any case. Once, the patient has failed (i.e., $Y(t) = 1$), the intervention stops or, equivalently, the interventions can be defined as requiring that the patient's exposure $a(t)$ remains unchanged after failure (i.e., $a_1(t) = a_1(t-1)$ and $a_2(t) = 0$).

We denote the decision rules that define the four interventions described above with four vectors of mappings $d_x = (d_x(0), \ldots, d_x(K))$ where $d_x(t) : (A(t-1), N(t-1), Y(t), I(t)) \mapsto (a_1(t), a_2(t))$ for $t = 0, ..., K$ and $x \in \mathcal{X} = \{7, 7.5, 8, 8.5\}$. To simplify notation below, the exposure regime through time $t$ defined by applying the sequence of decision rules $d_x$ with a patient's observed data $O$ is denoted by $d_x(\bar{V}(t)) = (d_x(0)(V(0)), \ldots, d_x(t)(V(t)))$, where the vector V(t) is defined as $V(t) = (A(t-1), N(t-1), Y(t), I(t))$. For a given A1c threshold $x$, we note that this exposure regime is not necessarily equal to the intervention of interest because, for example, we might have $A(0) \neq d_x(V(0))$.

In Section 2, we are interested in the counterfactual outcomes defined by the joint dynamic treatment-censoring intervention described above. Formally, these outcomes are defined by the NPSEM introduced in Section 1.4 and denoted by $Y_{d_x}(t+1)$. Contrasts of their distributions define the first set of causal parameters of interest in this report. More specifically, we first aim to estimate the difference in the counterfactual cumulative risk of failure at time $t_0$ between any two distinct regimes $d_{x_1}$ and $d_{x_2}$: $\psi^{d_{x_1}, d_{x_2}}(t_0 + 1) = P(Y_{d_{x_1}}(t_0 + 1) = 1) - P(Y_{d_{x_2}}(t_0 + 1) = 1)$.

## 2.2 Identifying assumptions

Identifiability of the above parameters with observational data relies on the no unmeasured confounding, and positivity assumptions.[11,9]

In the setting of longitudinal treatment interventions, the no unmeasured confounding assumption is also referred to as the sequential randomization assumption (SRA) which states that, conditional on the observed treatment and confounder history, each potential outcome of interest is independent of exposure status in each preceding time period. In settings where monitoring is performed in each period (i.e., when $\bar{N}(K - 1) = 1$), the sequential randomization assumption is stated as follows:

$$Y_{d_x}(t_0 + 1) \perp A(t) | \bar{L}(t), \bar{A}(t - 1), \tag{1}$$

for $t = 0, ..., t_0$, and $x \in \mathcal{X}$. This assumption is not testable in practice but its upholding requires that a sufficiently rich set of covariates are measured. Substantive subject-matter knowledge encoded in a causal diagrams may be used to inform the selection of relevant covariates (e.g., using the sequential back-door criterion[28]), but the adequacy of the selection then rests on the correct specification of the causal Directed Acyclic Graph (DAG) and in particular on the ability to assume that all common causes of any two nodes in the DAG are known.[25,28,29,30,31,32] For instance, the causal DAG at the top of Figure 1 represents commonly assumed causal relationships between observed treatment, covariates, and outcomes that support the upholding of the SRA in studies where covariates are monitored systematically at each time point (i.e., $I(t) = I^0(t)$). This DAG is the analog of the one presented in figure 1b of (33) but adapted to cohort studies with a time-to-event outcome. For simplicity, it is assumed that no covariate other than the A1c measurements and outcomes are collected (i.e., $Z(t)$ is nil), that follow-up spans only two time points (i.e., $K = 1$), and that right-censoring does not occur (i.e., $\bar{A}_2(K) = 0$ and the degenerate censoring nodes are thus omitted from the DAG just like the degenerate monitoring nodes $\bar{N}(K) = 1$). We note however that the arguments below hold in the more complex and realistic study scenarios with additional covariate nodes $Z(t)$, the occurrence of censoring events $A_2(t)$, and longer follow-up. In this DAG, each observed variable of the temporally ordered sequence $O$ is assumed to possibly affect all subsequent observed variables. In particular, we note that the first outcome measurement $Y(1)$, has an effect on the second outcome measurement $Y(2)$, the second covariate measurement $I(1)$, and the second treatment measurement $A_1(1)$ which reflects the convention that all variables become degenerate after failure. In addition to the observed variables, the DAG also includes two potentially unobserved time-varying covariates $U_1$ and $U_2$ (e.g., health-seeking behavior such as diet and physical activity in the TI study) that, as indicated by the dashed arrows, are risk factors for the outcomes $Y(t)$ and determinants of the covariates $I(t)$. The covariate $U_2$ may also be impacted by the prior observed covariate $I(0)$ and treatment $A_1(0)$. Because all backdoor paths from the treatment nodes $A_1(0)$ and $A_1(1)$ to the outcomes $Y(1)$ and $Y(2)$ are blocked by observed variables, the SRA holds. This DAG demonstrates that the SRA (1) will hold even in the presence of unmeasured risk factors for the outcome as long as the effects of these factors on treatment decisions are entirely mediated by covariates that are included in the observed data.

The study scenario just described can be extended to studies with non-systematic mon-itoring of covariates as shown by the DAG at the bottom of Figure 1. This second DAG is obtained by replacing the node $I(1) = I^0(1)$ in the first DAG with three new nodes: the indicator of A1c monitoring, $N(0)$, the latent A1c level $I^0(1)$, and the observed A1c level, $I(1)$. All parents of the node $I(1) = I^0(1)$ in the first DAG remain parents of the node $I^0(1)$ in the second DAG. Of the two children $A_1(1)$ and $Y(2)$ of the node $I(1) = I^0(1)$ in the first DAG, only the node $Y(2)$ remains a child of $I^0(1)$ in the second DAG.

The absence of an arrow from $I^0(1)$ to $A_1(1)$ encodes the assumption that the covariate $I^0(1)$ can only impact treatment decisions if it is observed (e.g., a clinician's decision to prescribe a new treatment cannot be influenced by the patient's A1c level if this level was not known to the clinician). Following the same rationale, both $N(0)$ and $I(1)$ become parents of the treatment node $A_1(1)$. We recall that the observed A1c measurement $I(1)$ is defined as $I^0(1)$ if both $N(0) = 1$ and $Y(1) = 0$. Otherwise, the covariate is defined by convention as the last observed value carried forward (LOVCF), i.e., $I(0)$. Therefore, the only parents of $I(1)$ in the second DAG are $I(0)$, $N(0)$, $Y(1)$, and $I^0(1)$. The node $N(0)$ is assumed to possibly impact directly both the latent covariate $I^0(1)$ and the outcomes $Y(1)$ and $Y(2)$.

In the second DAG, the following SRA holds since all backdoor paths between treatment nodes and outcomes are blocked by past observed outcomes, covariates, treatment, and monitoring nodes:

$$Y_{d_x}(t_0 + 1) \perp A(t) | \bar{L}(t), \bar{A}(t-1), \bar{N}(t-1) \tag{2}$$

In the second DAG, several backdoor paths from the treatment node $A_1(1)$ to the outcome $Y(2)$ are not blocked by the observed covariates $I(0)$, $I(1)$, treatment $A_1(0)$, and the outcome

$Y(1)$: e.g., $A_1(1)$–$N(0)$–$Y(2)$, $A_1(1)$–$N(0)$–$U_1$–$Y(2)$, and $A_1(0)$–$N(0)$–$I^0(1)$–$Y(2)$. The SRA as formulated by expression (1) may thus lead to biased estimation of the causal estimands $\psi^{d_{x_1}, d_{x_2}}$ in studies with non-systematic covariate monitoring if the monitoring variables $N(t)$ are not controlled for when applying methods based on the G-formula.[34] For example, bias could be expected from an IPW estimator constructed from propensity scores estimated with models that ignored the monitoring nodes. We note that the same concern also applies to causal estimands defined by static interventions on time-varying exposures. While the use of LOVCF to define covariate processes such as $\bar{I}(t)$ in studies with non-systematic covariate monitoring has become routine practice in the literature,[4,16,5,6] it is not clear that adjustment for the associated monitoring process $\bar{N}(t)$ has become common practice. The use of LOVCF without control for the process $\bar{N}(t)$ in studies with non-systematic covariate monitoring[35,36,37] can be viewed as an attempt to impute missing data on time-varying covariates such as $I^0(t)$ (i.e., when $N(t-1) = 0$) in order to emulate the observational study design represented by the first DAG in which covariates are collected at each time point (i.e., as if $\bar{N}(t) = 1$). The second DAG in Figure 1 demonstrates that unmeasured latent covariates such as $I^0(t)$ do not necessarily lead to biased effect estimation if the monitoring process $\bar{N}(t)$ is adjusted for in the analysis *and if it can be assumed that the effect of the latent variables on treatment decisions is entirely mediated by the observed covariate values $I(t)$ (an exclusion restriction assumption).* Imputation of $I^0(t)$ is then an unnecessary goal and, in fact, the second DAG shows that even if the imputation process is successful, residual confounding might be expected if the monitoring process $\bar{N}(t)$ is not controlled for (e.g., when constructing propensity scores for IPW estimation) because, for example, two of the backdoor paths from $A_1(1)$ to $Y(2)$ through $N(0)$ (via $U_1$ and $U_2$) would then remain open. Instead of viewing unmeasured values for covariates $I^0(t)$ as missing data, the second DAG of Figure 1 thus argues for an alternate approach in which the monitoring nodes $\bar{N}(t)$ are explicitly included in the observed data structure $O$ (as done in Section 1.3 or by including them in the vector of covariates $L(t)$). With this data structure, the SRA as originally formulated in (38) is equivalently expressed as (2) and its upholding can be supported by the exclusion restriction assumption above. We argue here that this assumption is often realistic for many covariates such as laboratory measurements in EHR-based studies and that controlling for monitoring nodes $\bar{N}(t)$ can then not only avoid the difficult goal of correctly imputing missing values for latent variables such as $I^0(t)$ but can also potentially prevent residual confounding from unblocked paths from $A_1(1)$ to outcomes $Y(2)$ through $N(0)$ in Figure 1.

In both studies with systematic and non-systematic covariate monitoring, identifiability of the causal estimands $\psi^{d_{x_1}, d_{x_2}}$ also hinges on upholding of a positivity assumption. It requires that for each regime $d_x$, in each period through $t_0$, patients have a positive probability of following that regime, conditional on having followed it up to that time point, not having failed yet, and any combination of their observed covariate, outcome, monitoring, and exposure history. We state the positivity assumption for studies in which covariate monitoring is not systematic:

$$P\Big(A(t) = d_x(t)(V(t)) \,\Big|\, \bar{L}(t), \bar{Y}(t) = 0, \bar{A}(t-1) = d_x(\bar{V}(t-1)), \bar{N}(t-1)\Big) > 0, \quad (3)$$

for $t = 0, \dots, t_0$.

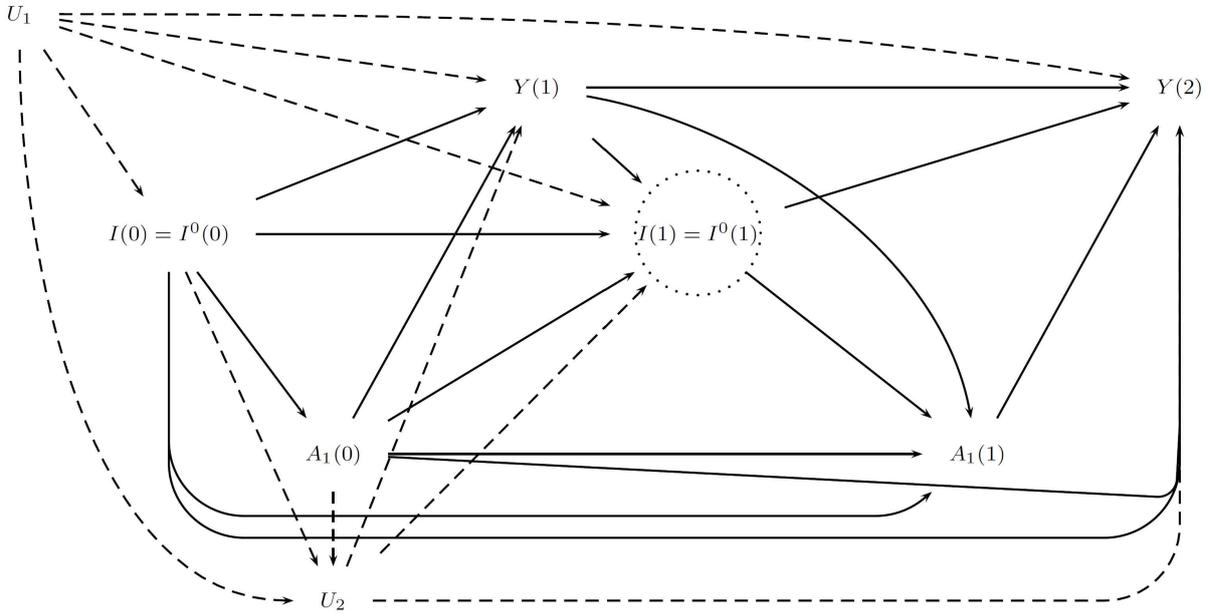

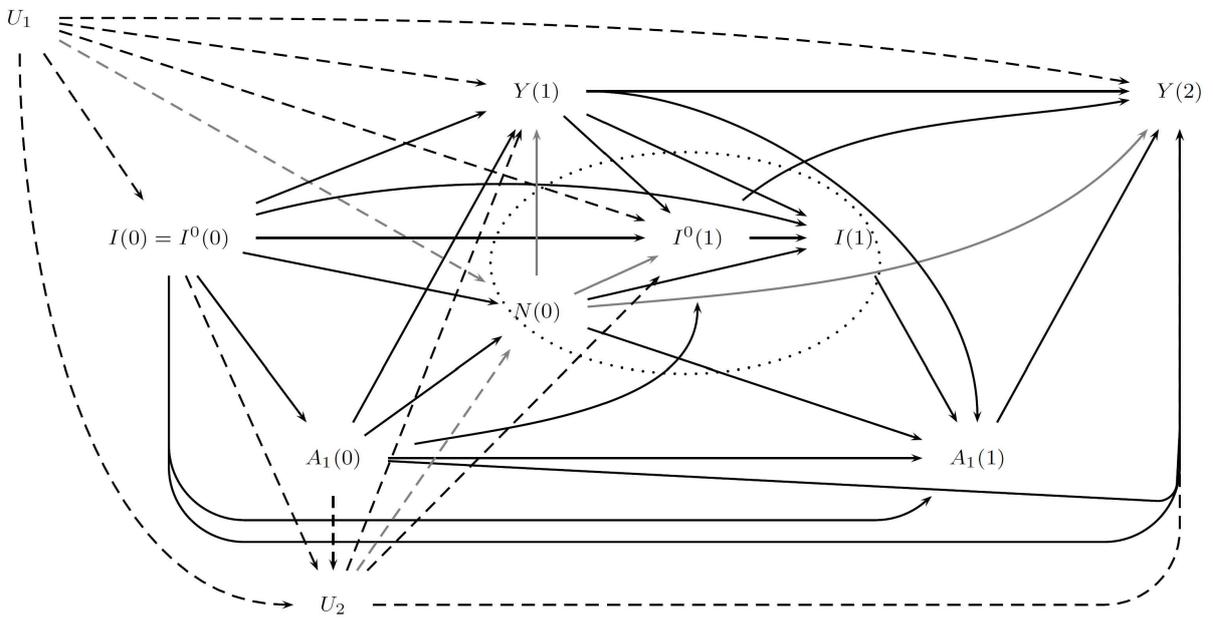

Figure 1: Examples of two Directed Acyclic Graphs (DAGs). The top DAG represents a commonly assumed scenario in cohort studies with time-varying covariates monitored at each time point (33, Figure 1b)

## 2.3 A hazard-based, bounded, IPW estimator

To estimate the causal risk difference, $\psi^{d_{x_1}, d_{x_2}}(t_0 + 1)$ under the SRA (2), we first separately estimate the two counterfactual cumulative risks $P(Y_{d_x}(t_0 + 1) = 1)$ with $x = x_1, x_2$ by IPW estimation. Second, we take the difference between the resulting estimates and base inference on the delta method.

More specifically, we evaluate each counterfactual risk using bounded IPW estimators of the discrete-time counterfactual hazards of failure. We selected this hazard-based IPW estimator because it is expected (as demonstrated in prior practical work[20]) to be more efficient than alternate, more direct IPW estimators of the cumulative risks.

The bounded IPW estimator of each counterfactual hazard is denoted by $P_n(Y_{d_x}(t + 1) = 1 | Y_{d_x}(t) = 0)$ and defined as a convex, linear combination of the outcomes at time $t + 1$ from all patients who did not experience the event before or at time $t$, and who followed the intervention of interest through time $t$:

$$\sum_{i=1}^{n} \frac{h_i(t)}{\sum_{i=1}^{n} h_i(t)} Y_i(t + 1), \tag{4}$$

where $h_i(t)$ is the inverse probability weight for individual $i$ defined by the function

$$h(t) = I(Y(t) = 0, \bar{A}(t) = d_x(\bar{V}(t))) \frac{\prod_{j=0}^{t} P_n'(A(j) = d_x(j)(\bar{V}(j)) | \bar{A}(j-1) = d_x(\bar{V}(j-1)))}{\prod_{j=0}^{t} P_n(A(j) | \bar{L}(j), \bar{A}(j-1), \bar{N}(j-1))}, \tag{5}$$

where $I(\cdot)$ denotes the indicator that event $\cdot$ occurs, $P_n(A(j) | \bar{L}(j), \bar{A}(j-1), \bar{N}(j-1))$ denotes an estimate of the conditional probability of the observed exposure at time $j$, given the past covariates, exposures, and monitoring decisions (i.e., the estimated propensity score for exposure when $A(j) = 1$), and $P_n'(A(j) = d_x(j)(\bar{V}(j)) | \bar{A}(j-1) = d_x(\bar{V}(j-1)))$ is a stabilizing factor, denoting the estimated probability of a patient following the regime $d_x$ of interest at time $j$ given that she followed the dynamic regime through time $j - 1$.

As noted elsewhere,[20] this estimator is equivalent to a stabilized IPW estimator of the coefficients of a *saturated* logistic dynamic marginal structural model (MSM) for the counterfactual hazards under the dynamic interventions $d_x$ with $x \in \mathcal{X}$ when these hazards are strictly between 0 and 1.[11,9] Such a dynamic MSM can be fitted with a standard weighted logistic regression using an expanded dataset[4] where each person-time observation is replicated for each dynamic regime a person follows at each time point. The weights to be used are the estimated inverse probability (IP) weights $h_i(t)$ defined above. We note that, when the MSM for the hazards is saturated, the use of stabilized IP weights $h_i(t)$ or their unstabilized version (i.e., (5) in which the probabilities $P_n'$ are replaced by 1) lead to the same IPW estimator (because these probabilities in the numerator and denominator of (4) cancel out), i.e., weight stabilization does not confer the gains in efficiency that originally motivated the stabilization of IP weights.[39] The estimated coefficients from the MSM fit are then transformed to obtain the estimate of the hazards, which are then mapped into an estimate of the counterfactual cumulative risk difference the following way:

$$\psi_n^{d_{x_1},d_{x_2}} = \prod_{t=0}^{t_0} \Big( 1 - P_n(Y_{d_{x_2}}(t+1) = 1 | Y_{d_{x_2}}(t) = 0) \Big) - \prod_{t=0}^{t_0} \Big( 1 - P_n(Y_{d_{x_1}}(t+1) = 1 | Y_{d_{x_1}}(t) = 0) \Big),$$

where $\psi_n^{d_{x_1},d_{x_2}}$ denotes the hazard-based IPW estimator of the causal estimand $\psi^{d_{x_1},d_{x_2}}$. For conciseness, we do not detail here a conservative estimator of its variance which is given in the Appendix of (20).

## 2.4 Implementation

In prior work,[20] the causal estimands $\psi^{d_{x_1}, d_{x_2}}$ were evaluated with the IPW estimator just described but its implementation implicitly relied on the SRA (1) instead of SRA (2). More specifically, the propensity scores that define the denominator of the weights (5) were estimated using models that ignored the monitoring variables $\bar{N}(j-1)$ and only considered the covariates $\bar{L}(j)$ and exposures $\bar{A}(j-1)$.

Motivated by the rationale on the upholding of the SRA in studies with non-systematic covariate monitoring outlined in Section 2.2, we revisit here the implementation of IPW estimation in the TI study to evaluate the practical impact of including versus excluding summary measures of the monitoring variables $\bar{N}(j-1)$ as new terms in the propensity score models that define the inverse probability weights (5).

In order to estimate the probability of the observed exposure given the observed past $P(A(j)|\bar{L}(j), \bar{A}(j-1), \bar{N}(j-1))$ referred to as the exposure assignment mechanism, we estimate separate propensity scores for the treatment and the censoring variables using various logistic models described in detail in prior work

We calculated the past frequency of A1c monitoring at each time point by dividing the number of prior periods in which A1C was monitored by the number of periods up to that time point, and we then created a categorical variable resulting in five categories approximately corresponding to the quintiles of the variable over all patients and time points: $[0, 0.4), [0.4, 0.5), [0.5, 0.7), [0.7, 1)$ and $[1]$, the last category describing those always monitored.

The categorical variable indicating whether the latest A1C measurement was monitored at the current period or carried forward from a previous period takes the value of 0 if the measurement was monitored at the current period (i.e., $N(j-1) = 1$), 1 if it was monitored at the previous period (i.e., $N(j-2) = 1$, $N(j-1) = 0$), 2 if it was monitored two periods earlier (i.e., $N(j-3) = 1$, $N(j-2) = N(j-1) = 0$), and so on, setting any levels higher than 4 to the value of 4. These categorical variables are then interacted with the latest measurement of A1C.

For covariates $Z(j)$ (e.g. LDL, blood pressure) that were not systematically monitored at each time point $j$, we followed a similar approach as the one used for $I(j)$, i.e., we used LOVCF to define the covariate when it was not monitored, and included the indicators of last value being carried forward in the vector of covariates $L(j)$ that define the main terms included in the logistic models for the propensity scores.

The estimates of the probabilities $P'_n$ that define the numerators of the IP weights (5) were derived non-parametrically using proportions of observed events. The resulting estimated stabilized IP weights were truncated[39] at the value 40.

## 2.5   Results

Figure 2 shows the estimated counterfactual survival curves over the first 2 years of follow-up before and after covariate adjustment with IPW for the four dynamic regimes studied. Figure 3 shows the adjusted point and interval estimates of the corresponding counterfactual cumulative risk differences. These results replicate the findings from prior analyses: Whereas unadjusted estimates of the cumulative risks provide little evidence of a protective effect of increasingly more aggressive treatment intensification strategies, adjusted estimates provide strong evidence that the risk of albuminuria development or progression almost always de-creases with the progressive decrease of the A1c threshold at which treatment intensification is initiated.

Tables 1 and 2 show the distribution of the stabilized and untruncated inverse probability weights among all person-time-regime observations who contribute an outcome to the effect estimates (i.e., an outcome that is assigned a nonzero IP weight) during the first 8 quarters of follow-up when the weights are estimated with logistic models that, respectively, ignore and include terms for the A1c monitoring variable $N(j-1)$ and their analogs for the other covariates $Z(j)$. We note a decrease in the number of large weights when derived with adjustment for monitoring (e.g., only 13 out of 651,250 IP weights are larger than the truncation level 40 compared to 28 IP weights>40 in the analysis without adjustment for monitoring). This change can be surprising given that, in our experience, the frequency of large weights often tend to increase as the number of covariates considered increases. The decrease in the frequency of large weights has little impact on the estimates of the survival curves (Figure 2) but leads, as expected,[39] to slightly more precise effect estimates as shown (Figure 3) by the estimated confidence intervals of the counterfactual risk differences over the first 8 periods of follow-up (2 years).

While results from these analyses do not demonstrate the potential for improved confounding adjustment when monitoring variables are included in the adjustment set, they do however illustrate that the inclusion of monitoring variables in the adjustment set does not impede appropriate confounding adjustment and that it can also improve estimation efficiency.

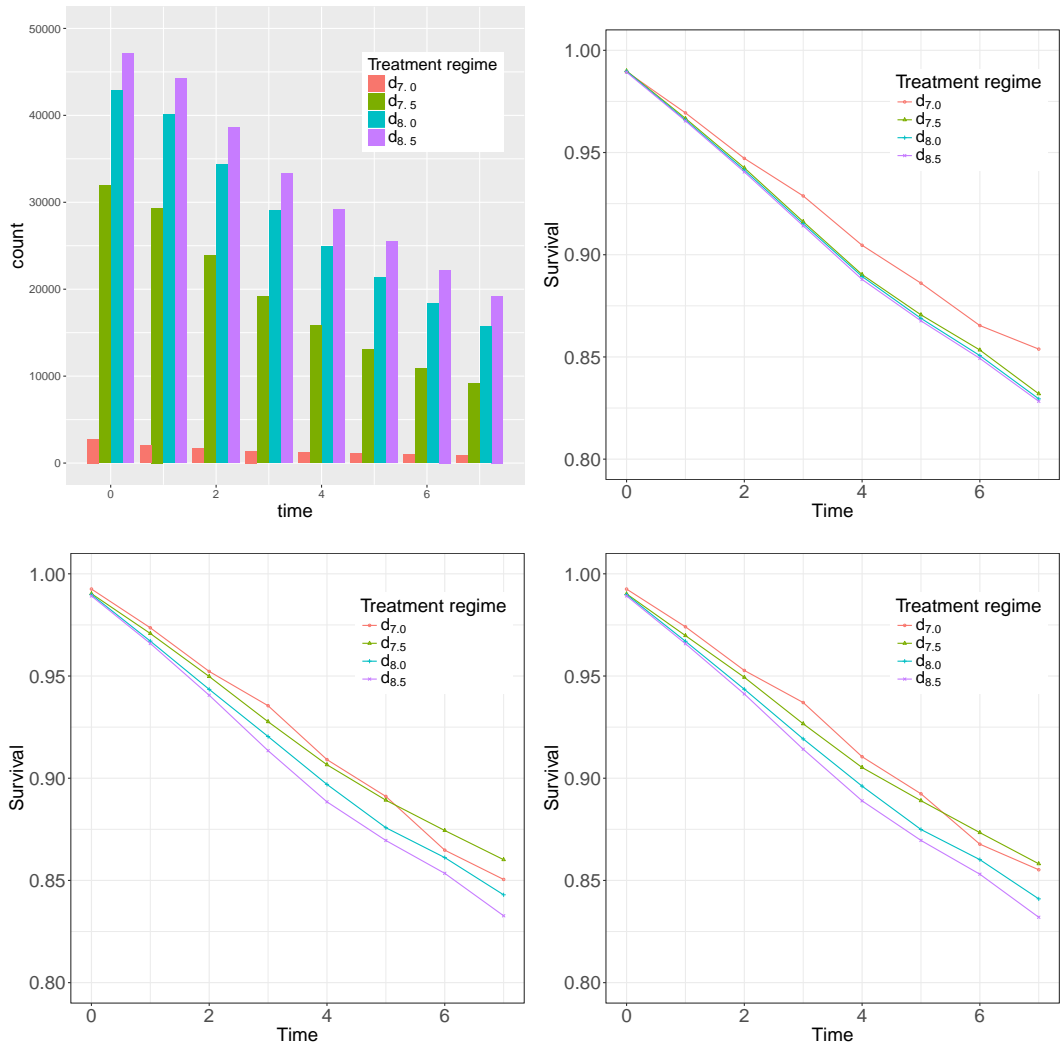

Figure 2: Crude (top right) and IPW estimates of the four counterfactual survival curves without (bottom left) and with (bottom right) adjustment for covariate monitoring in the propensity scores. The top left histogram displays the count of patients following each of the four dynamic treatment interventions over time.

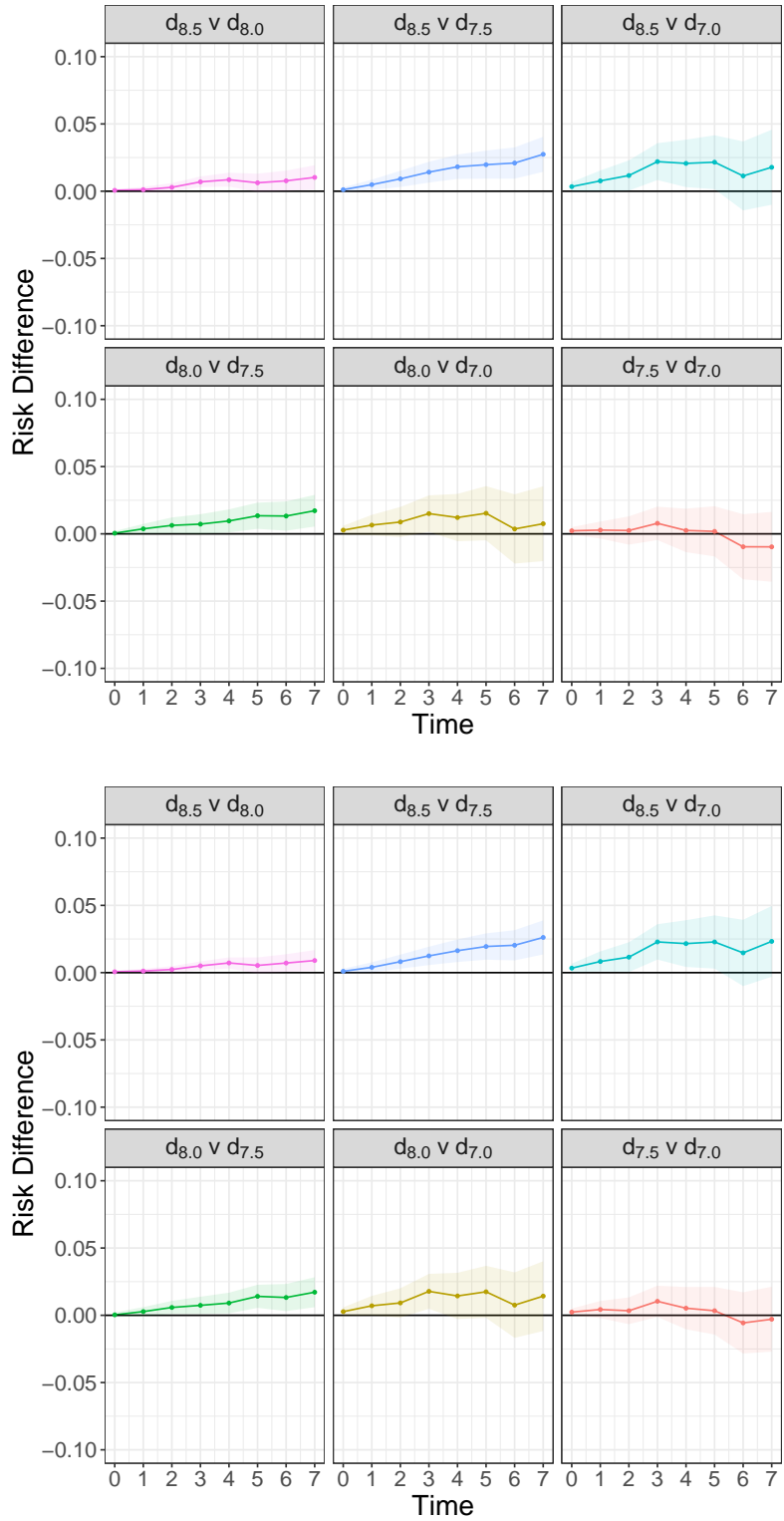

Figure 3: IPW estimates of counterfactual risk differences without (top) and with (bottom) adjustment for covariate monitoring in the propensity scores. The shaded areas represent the 95% confidence intervals.

Table 1: Distribution of stabilized inverse probability weights derived from propensity scores estimated based on logistic models without adjustment for covariate monitoring. Rule-person-time observations with a weight value of 0 are excluded.

| IPW | Frequency | % | Cumulative Frequency | Cumulative % |
|---|---|---|---|---|
| <0 | 0 | 0.00 | 0 | 0.00 |
| [0, 0.5[ | 95631 | 14.68 | 95631 | 14.68 |
| [0.5, 1[ | 496502 | 76.24 | 592133 | 90.92 |
| [1, 10[ | 53549 | 8.22 | 645682 | 99.15 |
| [10, 20[ | 4932 | 0.76 | 650614 | 99.90 |
| [20, 30[ | 540 | 0.08 | 651154 | 99.99 |
| [30, 40[ | 68 | 0.01 | 651222 | 100.00 |
| [40, 50[ | 21 | 0.00 | 651243 | 100.00 |
| [50, 100[ | 7 | 0.00 | 651250 | 100.00 |
| [100, 150[ | 0 | 0.00 | 651250 | 100.00 |
| ≥ 150 | 0 | 0.00 | 651250 | 100.00 |

Table 2: Distribution of stabilized inverse probability weights derived from propensity scores estimated based on logistic models with adjustment for covariate monitoring. Rule-person-time observations with a weight value of 0 are excluded.

| IPW | Frequency | % | Cumulative Frequency | Cumulative % |
|---|---|---|---|---|
| <0 | 0 | 0.00 | 0 | 0.00 |
| [0, 0.5[ | 106538 | 16.36 | 106538 | 16.36 |
| [0.5, 1[ | 481807 | 73.98 | 588345 | 90.34 |
| [1, 10[ | 59725 | 9.17 | 648070 | 99.51 |
| [10, 20[ | 2816 | 0.43 | 650886 | 99.94 |
| [20, 30[ | 294 | 0.05 | 651180 | 99.99 |
| [30, 40[ | 57 | 0.01 | 651237 | 100.00 |
| [40, 50[ | 8 | 0.00 | 651245 | 100.00 |
| [50, 100[ | 5 | 0.00 | 651250 | 100.00 |
| [100, 150[ | 0 | 0.00 | 651250 | 100.00 |
| ≥ 150 | 0 | 0.00 | 651250 | 100.00 |

# 3 Evaluation of joint dynamic treatment and static monitoring interventions

## 3.1 Causal estimands

Here we evaluate joint interventions that combine the previously defined dynamic intervention on treatment initiation with a static intervention on monitoring. This evaluation can be conducted either to determine the combination of a dynamic treatment strategy and A1c monitoring schedule that achieves the best health outcomes, to improve the generalizability of findings about the comparative effectiveness of the four dynamic treatment-censoring strategies to settings with specific A1c monitoring schedules, or to determine the frequency of A1c monitoring that achieve the best outcomes in settings where a specific dynamic treatment-censoring strategy is followed.

While any static monitoring regimes could be evaluated, for this study we focus on the following four regular monitoring schedules: monitor A1c every quarter (i.e., $\bar{n} = (1, 1, \ldots, 1)$), skip one quarter between A1c tests (i.e., $\bar{n} = (0, 1, 0, 1, \ldots)$), skip two quarters between A1c tests (i.e., $\bar{n} = (0, 0, 1, 0, 0, 1 \ldots)$), skip 3 quarters between A1c tests (i.e., $\bar{n} = (0, 0, 0, 1, 0, 0-, 0, 1, \ldots)$), where we recall that, by design, all patients have their A1c monitored at study entry (i.e., $N(-1) = 1$) and that the first element of each vector $\bar{n}$ denotes the decision made at time $t = 0$ on whether the patient's A1c will be monitored at time $t = 1$ (i.e., whether $I^0(1)$ is observed). The collection of these four static monitoring interventions is denoted by $\bar{\mathcal{N}}$.

We consider all sixteen possible joint interventions $(d_x, \bar{n})$ of the four previously defined dynamic treatment-censoring regimes $d_x$ with $x \in \mathcal{X}$ and the four static monitoring regimes $\bar{n} \in \bar{\mathcal{N}}$ defined above. For example, one possible joint regime of interest is "Monitor A1c levels every 3 months, and intensify diabetes treatment the first time A1c drifts above 7.5 %, and keep the patient on the intensified therapy thereafter." As before, each such intervention also requires that the patient remains uncensored. The counterfactual outcomes under these 16 interventions are defined by the NPSEM introduced in Section 1.4 and denoted by $Y_{d_x,\bar{n}}(t+1)$. Their distributions define the causal estimands of interest in this section:

$$\psi^{(d_{x_1},\bar{n}_1),(d_{x_2},\bar{n}_2)}(t_0+1) = P(Y_{d_{x_1},\bar{n}_1}(t_0+1)=1) - P(Y_{d_{x_2},\bar{n}_2}(t_0+1)=1). \qquad (6)$$

## 3.2   Identifying assumptions

In order to identify the causal parameter defined above, the SRA and positivity assumptions continue to be required but their expressions are modified as follows to reflect the change in the interventions of interest.

The SRA can now be expressed as:

$$Y_{d_x,\bar{n}}(t_0+1) \perp (A(t), N(t))|\bar{L}(t), \bar{A}(t-1), \bar{N}(t-1), \qquad (7)$$

for $t = 0, \ldots, t_0$, $x \in \mathcal{X}$, and $\bar{n} \in \mathcal{N}$. In practice, this assumption can be motivated by ensuring that all backdoor paths from $A_1(t)$, $A_2(t)$, and $N(t)$ to $Y(t_0+1)$ are blocked by prior measured variables (exposures or covariates) included in the observed data $O$. For instance, the SRA (7) will hold if the two gray dashed arrows from $U_1$ and $U_2$ to $N(0)$ are deleted from the DAG at the bottom of Figure 1. We note that it has been argued that the SRA (7) might often be less likely to hold than the SRA (2) in EHR-based studies because "data on the reason for a given visit are often not recorded for data analysis".[9] Although we cannot be sure that the SRA (7) holds in the EHR-based TI study, the stability in the point estimates from the secondary analyses with and without adjustment for the monitoring variables $N(t)$ presented in the prior section provides some support for the stronger assumption (7).

The positivity assumption can now be stated as follows:

$$\begin{cases} P\Big(A(t) = d_x(t)(V(t)) \,\Big|\, \bar{L}(t), \bar{Y}(t) = 0, \bar{A}(t-1) = d_x(\bar{V}(t-1)), \bar{N}(t-1) = \bar{n}(t-1)\Big) > 0, \\ P\Big(N(t) = n(t) \,\Big|\, \bar{L}(t), \bar{Y}(t) = 0, \bar{N}(t-1) = \bar{n}(t-1), \bar{A}(t) = d_x(\bar{V}(t))\Big) > 0, \end{cases}$$
$$(8)$$

for $t = 0, \ldots, t_0$, $x \in \mathcal{X}$, and $\bar{n} \in \bar{\mathcal{N}}$.

In particular, equation (8) requires that for all time periods and any combination of past covariates, there is a positive probability for each patient who previously followed the joint treatment-censoring-monitoring intervention to continue to follow the *monitoring* intervention of interest. Near-violations of this assumption can occur if certain covariates (e.g., change in A1c control) are strong determinants of monitoring decisions (e.g., the American Diabetes Association recommends that patients who recently changed treatments or whose A1c recently became out of control be monitored more frequently)

## 3.3   A hazard-based, bounded, IPW estimator

The bounded IPW estimator (4) is modified in the following way to derive an IPW estimator denoted by $P_n(Y_{d_x,\bar{n}}(t+1) = 1 | Y_{d_x,\bar{n}}(t) = 0)$ of the counterfactual hazards under each of the 16 joint dynamic treatment-censoring and static monitoring interventions described in this Section. The inverse probability weights $h_i(t)$ are now based on the joint conditional probability of receiving the observed exposure and the observed monitoring interventions.

More specifically, $h(t)$ is now defined as:

$$\frac{\prod_{j=0}^{t} P_n'\Big(A(j) = d_x(j)(\bar{V}(j)), N(j) = n(j) \mid \bar{A}(j-1) = d_x(\bar{V}(j-1)), \bar{N}(j-1) = \bar{n}(j-1)\Big)}{\prod_{j=0}^{t} P_n\Big(A(j) \mid \bar{L}(j), \bar{A}(j-1), \bar{N}(j-1)\Big) P_n\Big(N(j) \mid \bar{L}(j), \bar{A}(j), \bar{N}(j-1)\Big)}$$
$$\tag{9}$$

where $P_n(N(j)|\bar{L}(j), \bar{A}(j-1), \bar{N}(j-1))$ denotes an estimate of the conditional probability of the observed monitoring decision at time $j$, given the past covariates, exposures and monitoring decisions (i.e, an estimated propensity score for monitoring when $N(j) = 1$) and where

$$P_n'\Big(A(j) = d_x(j)(\bar{V}(j)), N(j) = n(j) \mid \bar{A}(j-1) = d_x(\bar{V}(j-1)), \bar{N}(j-1) = \bar{n}(j-1)\Big)$$

is a stabilizing factor denoting the estimated probability of a patient following both the dynamic treatment-censoring and static monitoring regimen $(d_x, \bar{n})$ at time $j$ given she followed both treatment-censoring and monitoring interventions $(d_x, \bar{n})$ through time $j - 1$.

The IPW estimators of the counterfactual hazards above are then mapped into an estimator $\psi_n^{(d_{x_1}, \bar{n}_1), (d_{x_2}, \bar{n}_2)}$ of the causal estimands (6) as follows:

$$\prod_{t=0}^{t_0} \Big(1 - P_n(Y_{d_{x_2}, \bar{n}_2}(t+1) = 1 | Y_{d_{x_2}, \bar{n}_2}(t) = 0)\Big) - \prod_{t=0}^{t_0} \Big(1 - P_n(Y_{d_{x_1}, \bar{n}_1}(t+1) = 1 | Y_{d_{x_1}, \bar{n}_1}(t) = 0)\Big)$$

and a conservative estimate of its variance is given by a straightforward extension of the results presented in the Appendix of.[20]

## 3.4 Implementation

In order to construct the weights defined in (9), beyond estimating the propensity scores for the treatment and censoring variables using the logistic regressions described in Section 2.4, we now also estimate the probability of A1c monitoring at each time point $t$ given the treatment at time $t$ and the observed past, i.e., $P(N(t) = 1|\bar{L}(t), \bar{A}(t), \bar{N}(t-1))$ by fitting a single logistic regression model with data pooled over all time periods. The model includes a term for time $t$, the treatment at time $t$ ($A_1(t)$) and each of the the same covariates ($L(t)$) and summary measures of their histories through period $t$ used in the propensity score model for the exposure variable described in Section 2.4, among them the categorical variables indicating the past frequency of A1C monitoring, the categorical variable indicating whether the latest A1c measurement was monitored at the current period or carried forward from a previous period, and their interactions with the latest A1c value ($I(t)$). In addition, we included terms for the categorical variables that encode a discretized version of the observed A1c $I(t)$ corresponding with the A1c intervals $< 7\%$; $]7\%;7.5\%]$; $]7.5\%;8\%]$; $]8\%;8.5\%]$; $> 8.5\%$. For all covariates $Z(t)$ (e.g. LDL, blood pressure) that were not systematically monitored at each time point $t$, we also used LOVCF to define the covariate when it was not monitored, and included the indicators of last value being carried forward in the vector of covariates $L(t)$ that define the main terms included in the logistic models for the propensity scores. The estimates of the probabilities $P'_n$ that define the numerators of the IP weights (9) were derived non-parametrically using proportions of observed events. The estimated stabilized IP weights were truncated at the value 40.

## 3.5 Results

Figure 4 shows the numbers of patients following each of 16 joint treatment-censoring-monitoring interventions for the first 8 periods and can be compared to their analogs without monitoring interventions in Figure 2 (top left panel). These results confirm that the counts of patients following any of the joint interventions are much lower in general than their analogs without monitoring interventions. As shown in Figure 2, the information to evaluate the dynamic treatment-censoring intervention $d_{7.0}$ is relatively limited compared to the other three dynamic treatment-censoring interventions. When combining the intervention $d_{7.0}$ with any of the four monitoring interventions, Figure 4 (top left panel) indicates that there is very little information left to evaluate such joint interventions.

Whichever the dynamic intervention, Figure 4 also indicates that the joint interventions that require monitoring at each period (denoted by $\bar{n}_0$) are followed by significantly fewer patients than those that require consecutive A1C tests to be separated by 1, 2 or 3 periods (denoted by $\bar{n}_1, \bar{n}_2$, and $\bar{n}_3$, respectively). Whichever the dynamic intervention, Figure 4 also indicates that the counts of patients who follow any of the joint interventions indexed by the last three monitoring patterns are very similar at each time point, and as expected, the counts are in fact identical at time 0 (because all three monitoring patterns require that $n(0) = 0$) and remain identical for $\bar{n}_2$ and $\bar{n}_3$ at time 1 because both continue to require $n(1) = 0$ (unlike $\bar{n}_1$ for which $n_1(1) = 1$). Starting at time 2, the counts for all joint interventions indexed by these three monitoring patterns are systematically different but very close in magnitude. Despite the similarity in available information to evaluate the joint interventions defined by $\bar{n}_1, \bar{n}_2$, and $\bar{n}_3$, we note that the number of patients who follow joint interventions indexed by the monitoring pattern $\bar{n}_1$ is almost always slightly larger than all other joint interventions indexed by the other three monitoring patterns. In addition, the counts of patients who follow each of the 9 joint interventions $(d_x, \bar{n}_j)$ for $x \neq 7.0$ and $j \neq 0$ at time 0 is consistently lower by about 8,000 patients compared to their analogs shown on Figure 2. This drop in the counts of patients following these 9 joint interventions becomes more drastic at time 1 and reaches at least 20,000 patients. Beyond the 6th period (a year and a half), none of the joint regimes are followed by more than a few patients.

We contrast the counterfactual survival curves denoted by the 16 joint treatment censoring-monitoring interventions, first by comparing the four dynamic treatment-censoring interventions when they are all combined with the same monitoring intervention (Figures 5 and 6), and second by comparing the four monitoring interventions when they are all combined with the same dynamic treatment-censoring intervention (Figures 7 and 8). The decrease in available information described earlier when combining the four dynamic treatment-censoring strategies with a particular monitoring pattern explains the large increase in the standard errors of the IPW risk difference estimators (i.e., the width of confidence intervals) displayed in Figure 6 compared to those displayed at the bottom of Figure 3. Despite this large reduction in precision, all plots except for the bottom right one on Figure 5 show a similar separation and ordering of the survival curves compared to those obtained without monitoring interventions and displayed in the bottom right plot on Figure 2. In other words, these results also suggest that the risk of onset or progression of albumbinuria decreases with the decrease of the A1c threshold at which an intensified treatment is initiated. We note that the increase in estimation variability also likely explains why the survival curves in Figure 5 are erratic compared to those in 2. The strongest statistical evidence to support a protective effect of treatment intensification at a lower A1c threshold is obtained for the monitoring intervention $\bar{n}_1$ as shown on the top right plot of Figure 6. When contrasting joint interventions defined by different monitoring interventions but the same dynamic treatment-censoring intervention on Figure 7, results suggest a beneficial effect of more frequent monitoring on the risk of onset or progression of albuminuria. Despite the relatively consistent trend in the point estimates of the risk difference on Figure 8, the wide confidence intervals do not provide strong statistical evidence of a protective effect of more frequent A1c monitoring.

Table 3 displays the distribution of the stabilized and untruncated IP weights for person-time-regime outcomes during the first 8 quarters of follow-up with nonzero weight values. We note that the distribution of the IP weights is shifted right compared to that for regimes without monitoring interventions displayed in Table 2 and, in particular, the number of weights greater than or equal to 40 and 150 is now 110 and 13, respectively, out of 597,111 observations contributing to the analysis of joint interventions compared to 13 and 0 out of 651,250 in the analysis without monitoring intervention. This shift and increase in the number of large weights confirm theoretical concerns over increased near-violations of the positivity assumption (8) when evaluating dynamic treatment-censoring interventions combined with monitoring interventions. Table 4 indicates that almost all weights greater than 40 and 150 are assigned to observations contributing to the evaluation of joint interventions that require continuous A1c monitoring. Based on the difference between the distributions of the weights in Tables 3 and 4 and the histograms in Figure 4, we conjecture that the observed increase in near-violations of the positivity assumption is likely resulting from the large reduction in the number of patients following each joint intervention (i.e., it could be avoided with increased sample sizes) but note that it may also be indicative of the existence of covariate events that are strong determinants of monitoring decisions (i.e., structural violations that would persist with increased sample sizes).

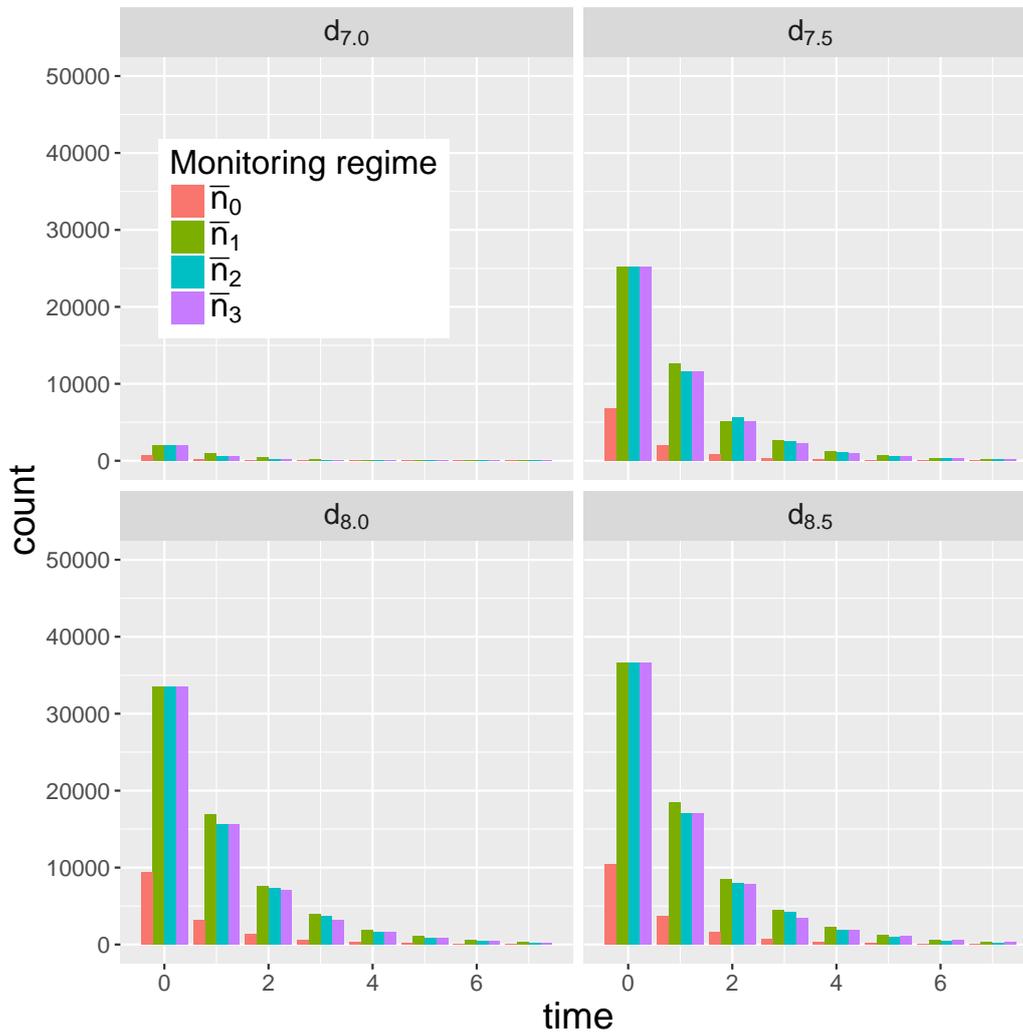

Figure 4: Counts of patients following each of the 16 joint treatment-censoring and monitoring interventions over time. Counts are organized by dynamic treatment-censoring interventions. The static monitoring interventions require that two consecutive A1c tests always be separated by 0, 1, 2, and 3 periods and are denoted by $\bar{n}_0$, $\bar{n}_1$, $\bar{n}_2$, and $\bar{n}_3$, respectively.

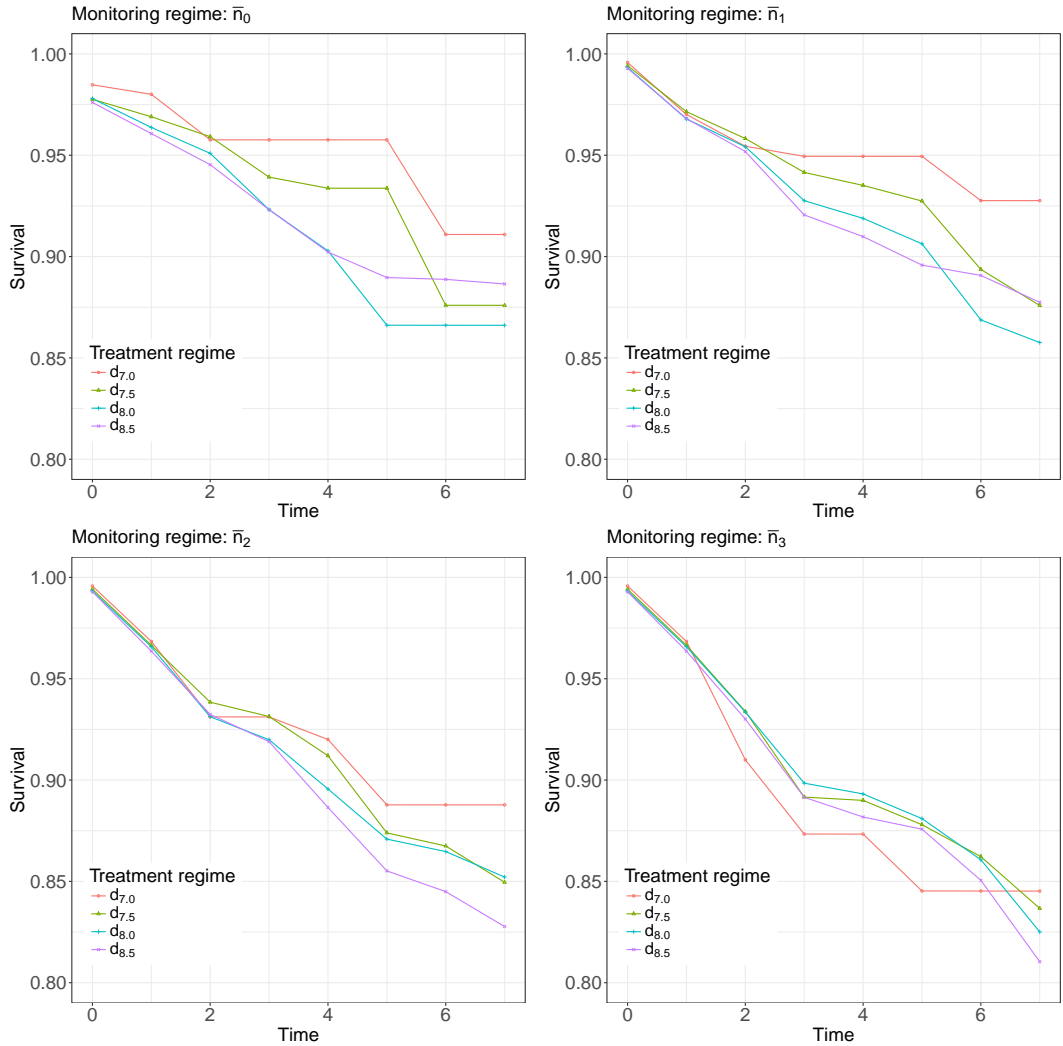

Figure 5: IPW estimates of the counterfactual survival curves under the 16 joint dynamic treatment-censoring-monitoring interventions. Each plot contrasts the four counterfactual survival curves defined by four distinct dynamic treatment-censoring interventions and a single static monitoring intervention. Clockwise starting with the top left plot, the static monitoring intervention requires that two consecutive A1c tests always be separated by 0, 1, 3, and 2 periods and is denoted by $\bar{n}_0$, $\bar{n}_1$, $\bar{n}_3$, and $\bar{n}_2$, respectively, in the caption of each plot.

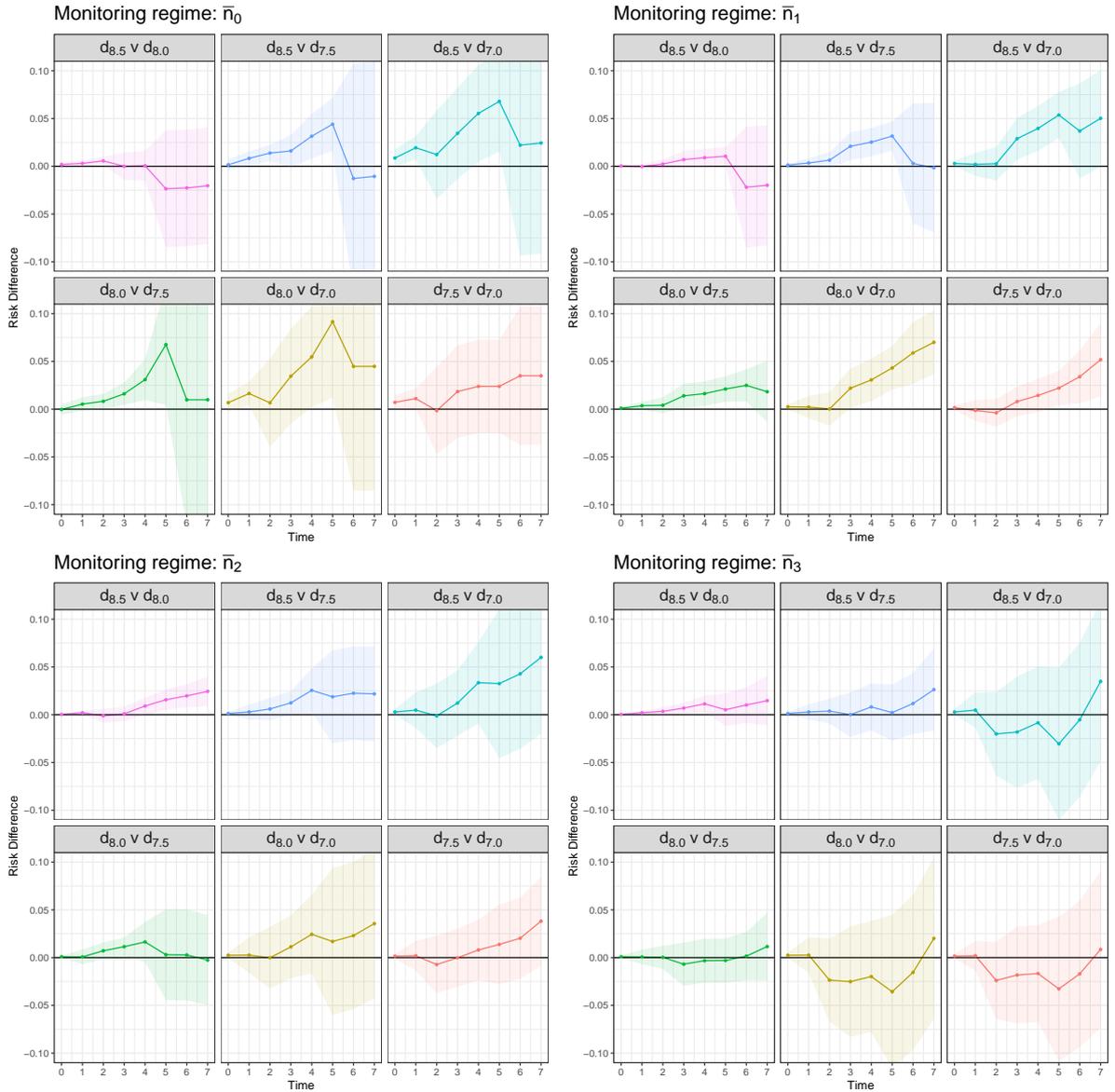

Figure 6: IPW estimates of counterfactual risk differences between dynamic treatment-censoring regimes combined with a given static monitoring intervention. The shaded areas represent the 95% confidence intervals. Clockwise, starting with the top left plot, the static monitoring intervention requires that two consecutive A1c tests always be separated by 0, 1, 3, and 2 periods and is denoted by $\bar{n}_0$, $\bar{n}_1$, $\bar{n}_3$, and $\bar{n}_2$, respectively, in the caption of each plot.

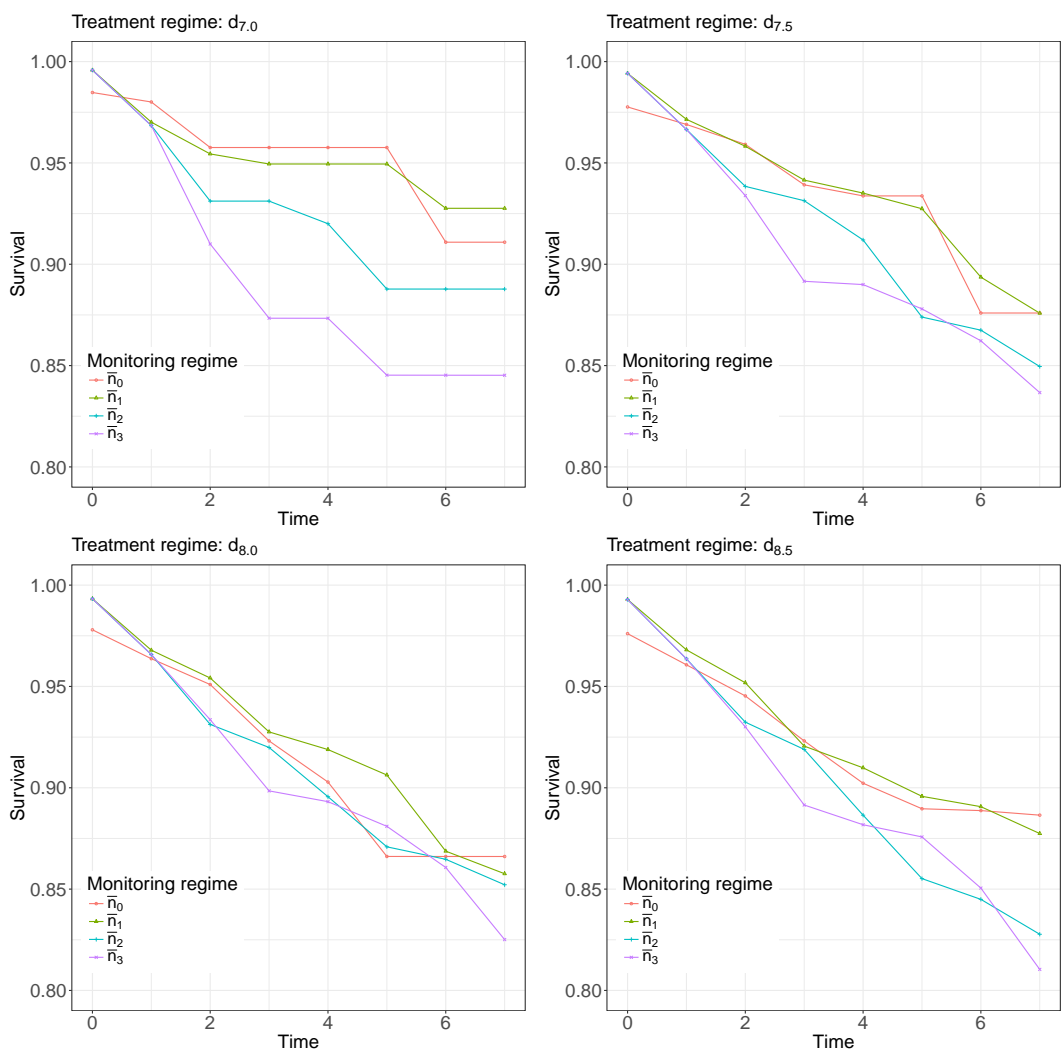

Figure 7: IPW estimates of the counterfactual survival curves under the 16 joint dynamic treatment-censoring and monitoring interventions. Each plot contrasts the four counterfactual survival curves defined by four distinct static monitoring interventions and a single dynamic treatment-censoring intervention. The static monitoring regimes contrasted on each plot require that two consecutive A1c tests always be separated by 0, 1, 2, and 3 periods and are denoted by $\bar{n}_0$, $\bar{n}_1$, $\bar{n}_2$, and $\bar{n}_3$, respectively.

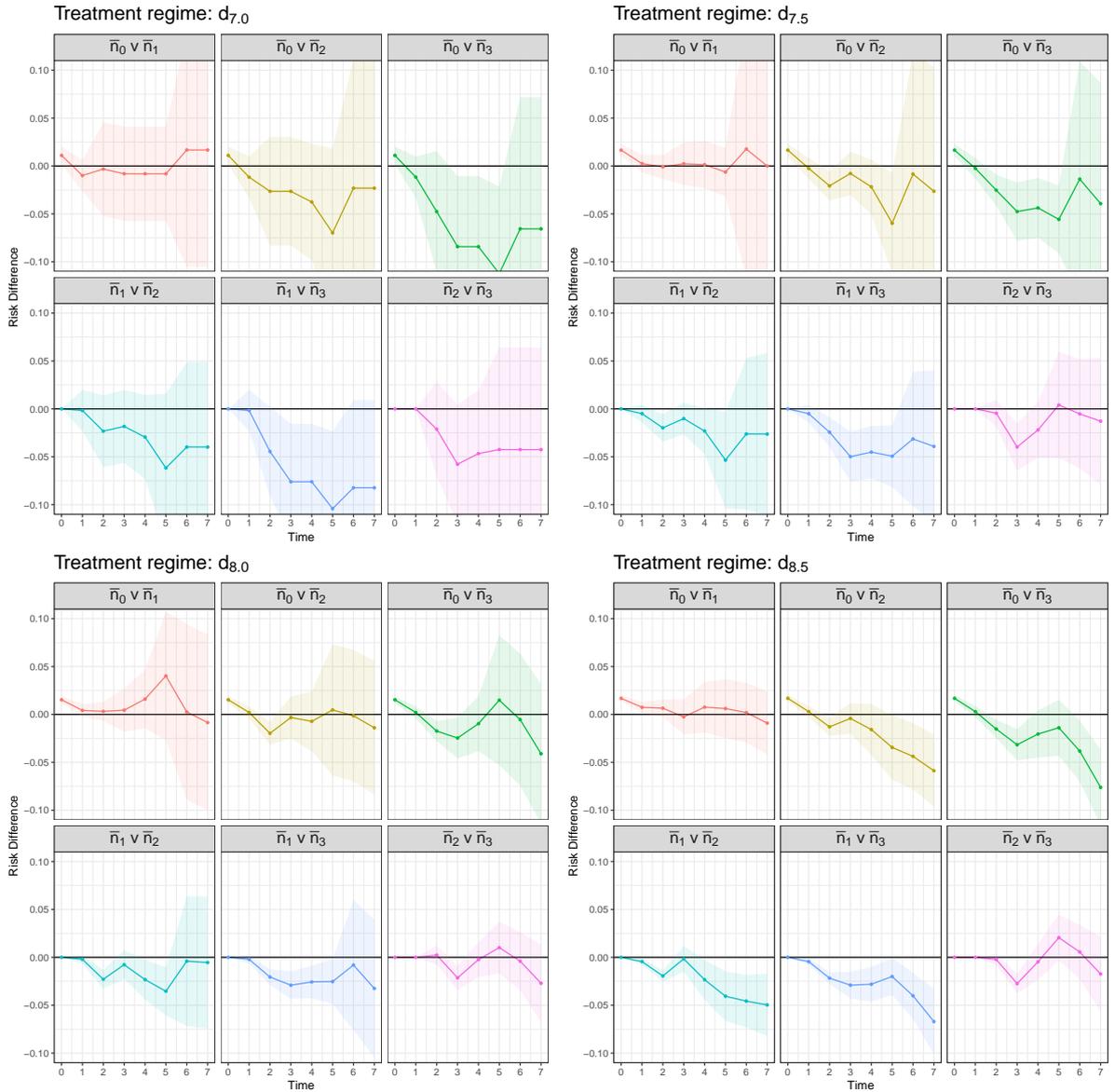

Figure 8: IPW estimates of counterfactual risk differences between static monitoring regimes combined with a given dynamic treatment-censoring intervention. The shaded areas represent the 95% confidence intervals. The four static monitoring interventions contrasted in each plot require that two consecutive A1c tests always be separated by 0, 1, 2, and 3 periods and are denoted by $\bar{n}_0$, $\bar{n}_1$, $\bar{n}_2$, and $\bar{n}_3$, respectively.

Table 3: Distribution of stabilized inverse probability weights for evaluating causal estimand (6). Rule-person-time observations with a weight value of 0 are excluded.

| IPW | Frequency | % | Cumulative Frequency | Cumulative % |
|---|---|---|---|---|
| $<0$ | 0 | 0.00 | 0 | 0.00 |
| $[0, 0.5[$ | 24829 | 4.16 | 24829 | 4.16 |
| $[0.5, 1[$ | 364741 | 61.08 | 389570 | 65.24 |
| $[1, 10[$ | 204120 | 34.18 | 593690 | 99.43 |
| $[10, 20[$ | 2821 | 0.47 | 596511 | 99.90 |
| $[20, 30[$ | 392 | 0.07 | 596903 | 99.97 |
| $[30, 40[$ | 98 | 0.02 | 597001 | 99.98 |
| $[40, 50[$ | 34 | 0.01 | 597035 | 99.99 |
| $[50, 100[$ | 56 | 0.01 | 597091 | 100.00 |
| $[100, 150[$ | 7 | 0.00 | 597098 | 100.00 |
| $\geq 150$ | 13 | 0.00 | 597111 | 100.00 |

Table 4: Distribution of the subset of stabilized inverse probability weights for evaluating causal estimand (6) when $\bar{n}_1$ and $\bar{n}_2$ do not enforce continuous A1c monitoring. Rule-person-time observations with a weight value of 0 are excluded.

| IPW | Frequency | % | Cumulative Frequency | Cumulative % |
|---|---|---|---|---|
| $<0$ | 0 | 0.00 | 0 | 0.00 |
| $[0, 0.5[$ | 19804 | 3.58 | 19804 | 3.58 |
| $[0.5, 1[$ | 334828 | 60.51 | 354632 | 64.09 |
| $[1, 10[$ | 195915 | 35.40 | 550547 | 99.49 |
| $[10, 20[$ | 2461 | 0.44 | 553008 | 99.93 |
| $[20, 30[$ | 292 | 0.05 | 553300 | 99.99 |
| $[30, 40[$ | 55 | 0.01 | 553355 | 100.00 |
| $[40, 50[$ | 11 | 0.00 | 553366 | 100.00 |
| $[50, 100[$ | 3 | 0.00 | 553369 | 100.00 |
| $[100, 150[$ | 0 | 0.00 | 553369 | 100.00 |
| $\geq 150$ | 0 | 0.00 | 553369 | 100.00 |

# 4 Evaluation of joint dynamic treatment-censoring and static monitoring interventions under the no direct effect assumption

## 4.1 Causal estimands

Under the NDE assumption detailed below, the causal parameters (6) in the prior Section were shown by[15] to equal the following causal estimands that are of interest here:

$$\psi^{(d^*_{x_1}, g^*_1),(d^*_{x_2}, g^*_2)}(t_0+1) = P(Y_{d^*_{x_1}, g^*_1}(t_0+1) = 1) - P(Y_{d^*_{x_2}, g^*_2}(t_0+1) = 1), \qquad (10)$$

where each $d^*_{x_j}$ with $x_j \in \mathcal{X}$ is a modified version of the dynamic treatment-censoring intervention $d_{x_j}$ and where each $g^*_j$ is a static monitoring intervention on a subset of the monitoring process. Both interventions $d^*_{x_j}$ and $g^*_j$ are defined as follows based on the same static monitoring intervention $\bar{n}_j \in \mathcal{N}$. Instead of requiring a patient to follow the pre-specified monitoring intervention $\bar{n}_j$ at every time point, the static monitoring intervention $g^*_j$ requires patients to be monitored at least at the pre-specified set of time points $t$ when $n_j(t) = 1$, but does not constrain monitoring decisions at other time points, i.e., it allows for the monitoring process to take its natural course for the remaining time points $t$ when $n_j(t) = 0$. The modified dynamic treatment-censoring intervention $d^*_{x_j}$ differs from the previously defined decision rule $d_{x_j}$ in that it only uses the past A1c measurements which are observed both under the monitoring regime $\bar{n}_j$, and under the actual observed monitoring process $\bar{N}(j)$, i.e., $d^*_{x_j}(t)(V(t)) = d_{x_j}(t)(V^*(t))$ where we recall that $V(t) = (A(t-1), N(t-1), Y(t), I(t))$ and where we define $V^*(t) = (A(t-1), n_j(t-1)N(t-1), Y(t), n_j(t-1)I(t))$. Thus, the modified dynamic treatment-censoring interventions $d^*_{x_j}$ corresponds to the original dynamic treatment-censoring intervention $d_{x_j}$ except that it requires that any additional A1c measurements collected beyond the A1c measurements required under intervention $\bar{n}_j$ be ignored when applying the decision rule $d_{x_j}$.

## 4.2 Identifying assumptions

While the causal estimands (10) can be of interest in their own rights, we focus our discussions on their use as an indirect approach to the evaluation of the causal estimands (6) introduced in the prior Section. We note that the identifiability of the causal estimands (10) relies only on the SRA and positivity assumption detailed below. However, if the causal estimands (10) are evaluated as an indirect mean to estimating the causal estimands (6), an additional identifiability assumption referred to as the No Direct Effect assumption is required. This assumption is formalized by the equality of counterfactual covariates under two sets of static interventions introduced in:[9]

$$L_{\bar{a},\bar{n}}^0(t) = L_{\bar{a}}^0(t), \tag{11}$$

where $L^0(t) = (Y(t), Z(t), I^0(t))$. In the NPSEM framework, it was shown by (15) that this assumption is implied by an exclusion restriction assumption that requires that all directed paths from nodes $N(t)$ to subsequent covariates $Y(j)$, $Z(j)$, and $I^0(j)$ be intercepted by treatment or censoring nodes $A(t)$ for $j > t$. In the DAG at the bottom of Figure 1, this assumption is encoded by the exclusion of the three solid gray arrows from $N(0)$ to $Y(1)$, $I^0(1)$, and $Y(2)$. Under the NDE assumption, the counterfactual outcomes defined by the joint treatment, censoring, and monitoring intervention $(d_{x_j}^*, g_j^*)$ was shown by[15] to equal the counterfactual outcomes of the joint intervention $(d_{x_j}, \bar{n}_j)$. The causal estimands (6) and (10) then become equivalent, i.e., $\psi^{(d_{x_1}, \bar{n}_1),(d_{x_2}, \bar{n}_2)}(t_0 + 1) = \psi^{(d_{x_1}, g_1^*),(d_{x_2}, g_2^*)}(t_0 + 1)$.

Consequently, the IPW estimator of the causal estimands (10) that we describe in the next section can be used as an indirect estimator of the causal estimands (6) when the NDE assumption holds. The potential advantage of this indirect IPW estimator over the direct IPW estimator described in Section 3.3 can be brought to light by comparing the SRA and positivity assumption required to identify estimands (10) with their analogs required to identify estimands (6).

The SRA required for identification of the causal estimand (10) can be expressed as:

$$
\begin{cases}
Y_{d^*_{x_j}, g^*_j}(t_0 + 1) \perp (A(t), N(t)) | \bar{L}(t), \bar{A}(t-1), \bar{N}(t-1) \text{ for all } t \text{ such that } n_j(t) = 1 \\
Y_{d^*_{x_j}, g^*_j}(t_0 + 1) \perp A(t) | \bar{L}(t), \bar{A}(t-1), \bar{N}(t-1) \text{ for all } t \text{ such that } n_j(t) = 0,
\end{cases}
\tag{12}
$$

with $t = 0, \ldots, t_0$. We note that this assumption is weaker than SRA (7) in the sense that it holds even when some backdoor paths from $N(t)$ to $Y(t_0 + 1)$ are not blocked by variables in $O$ collected before $N(t)$ and as long as such open backdoor paths only occur at time points $t$ when $n_j(t) = 0$. However, this change in the SRA requirement is not likely to be of much practical value because we do not expect it to result in the selection of different covariate sets for confounding and selection bias adjustment in practice. In other words, we expect that most arguments in favor of or against the upholding of SRA (7) will typically also apply to the upholding of the SRA (12) and vice versa.

The positivity assumption for identification of the causal estimand (10) can be stated as:

$$
\begin{cases}
P\Big(A(t) = d^*_{x_j}(t)(V(t)) \,\Big|\, \bar{L}(t), \bar{Y}(t) = 0, \bar{A}(t-1) = d^*_{x_j}(\bar{V}(t-1)), \\
\quad \bar{N}(t-1) = g^*_j(\bar{N}(t-1))\Big) > 0 \\
\quad \text{for } t = 0, \ldots, t_0 \text{ and,} \\
P\Big(N(t) = 1 \,\Big|\, \bar{L}(t), \bar{Y}(t) = 0, \bar{N}(t-1) = g^*_j(\bar{N}(t-1)), \bar{A}(t) = d^*_{x_j}(\bar{V}(t))\Big) > 0 \\
\quad \text{for } t = 0, \ldots, t_0 \text{ such that } n_j(t) = 1,
\end{cases}
\tag{13}
$$

where $g^*_j(\bar{N}(t)) = \big(g^*_j(0)(N(0)), \ldots, g^*_j(t)(N(t))\big)$ is defined by the mappings $g^*_j(k) : N(k) \mapsto N(k)^{1-n_j(k)}$ for $k = 0, \ldots, K$ as the sequence of monitoring decisions through time $t$ that is consistent with following intervention $g^*_j$ and compatible with the subset of observed monitoring decisions that are not constrained by intervention $g^*_j$.

We note that the positivity assumption (13) requires that the A1c of a patient who previously followed the intervention $(d^*_{x_j}, g^*_j)$ be possibly monitored at only the time points $t \leq t_0$ when $n_j(t) = 1$, whichever the patient's covariate and monitoring history. This is in contrast to the positivity assumption (8) that requires that a patient who previously followed the intervention $(d_{x_j}, \bar{n}_j)$ have its A1c monitored at all time points $t \leq t_0$ according to the static intervention $n_j(t)$, whichever the patient's covariate history. It can thus be expected that the positivity assumption (13) will often be more likely to hold in practice than its analog (8). For example, if $\bar{n}_j = (0, 0, 1, 0, 0, 1, \ldots)$, the positivity assumption (13) only requires that each patient can possibly be monitored every third quarter and does not place any constraint on the monitoring decisions between these yearly monitoring events, while the positivity assumption (8) requires that each patient can possibly experience the exact sequence of monitoring decisions $\bar{n}_j$ throughout follow-up.

The practical consequences of this weakening of the positivity assumption are expected improvements in IPW estimation performance, both in terms of finite sample bias and precision because of the likely more stable IP weights (e.g., fewer extreme weight values or more compact weight distribution), and the increased number of person-time observations with nonzero weight values, i.e., the increase in the number of patients whose observed exposure and monitoring history is consistent with following the joint treatment-censoring-monitoring intervention at each time point. However, we note that this potential for improved practical performance in the estimation of the causal estimand (10) over that of (6) might be offset by the fact that the positivity assumption for the exposure process in (13) can conceivably be more likely violated in practice than that in (8).

Indeed, the specification of the dynamic interventions $d_{x_j}$ that are evaluated in practice often integrates considerations about the occurrence of intermediate events (e.g., contraindications) that should preclude certain treatment decisions to occur.[42] In fact, it is such a consideration that originally motivated[16] the evaluation of the four dynamic interventions studied in this paper over static alternatives in the TI study (e.g., start an intensified treatment 1 years versus 2 years after study entry). Specifically, the expectation that most patients would not remain unexposed to an intensified therapy if their A1c drifted above a high threshold (>8.5%) raised concerns over the violation of the positivity assumption that would be required for identifying the effects of static interventions. This same consideration raises concerns here over a possible violation of the positivity assumption requirements for the exposure process in (13) because the treatment decision at time $t$ according to rule $d_{x_j}^*$ would require, for example, that a patient remain unexposed to an intensified therapy even if an A1c measurement collected at time $t$ reached a very high level and that simply because this measurement $I(t)$ happened to be collected at a time point $t$ when $\bar{n}_j(t) = 0$. Later in this section, we examine the trade-off between improved IP weight stability resulting from interventions on fewer monitoring nodes with $g_j^*$ and worsened IP weight stability resulting from poorer adherence to decision rule $d_{x_j}^*$ by comparing the practical performance of the IPW estimator introduced next to that of the IPW estimator from Section 3.3 for estimating the causal estimand (6) under the NDE assumption.

## 4.3   A hazard-based, bounded, IPW estimator

The bounded IPTW estimator introduced in Section 3.3 for evaluating the counterfactual hazards under the 16 joint dynamic treatment-censoring and static monitoring monitoring interventions $(d_x, \bar{n})$ is modified in the following way to derive an IPW estimator denoted by $P_n(Y_{d_{x_j}^*, g_j^*}(t+1) = 1 | Y_{d_{x_j}^*, g_j^*}(t) = 0)$ of the counterfactual hazards under each of the 16 interventions $(d_{x_j}^*, g_j^*)$. The inverse probability weights $h_i(t)$ are now defined by the modified function $h(t)$:

$$\frac{\prod_{k=0}^{t} P_n' \Big( A(k) = d_{x_j}^*(k)(\bar{V}(k)), N(k) = g_j^*(k)(N(k)) \mid \bar{A}(k-1) = d_{x_j}^*(\bar{V}(k-1)), \bar{N}(k-1) = g_j^*(\bar{N}(k-1)) \Big)}{\prod_{k=0}^{t} P_n \Big( A(k) \mid \bar{L}(k), \bar{A}(k-1), \bar{N}(k-1) \Big) \prod_{k=0,\ldots,t : n_j(k)=1} P_n \Big( N(k) \mid \bar{L}(k), \bar{A}(k), \bar{N}(k-1) \Big)}, \quad (14)$$

$$I \Big( Y(t) = 0, \bar{A}(t) = d_x(\bar{V}(t)), \bar{N}(t) = g_j^*(\bar{N}(t)) \Big)$$

where

$$P_n' \Big( A(k) = d_{x_j}^*(k)(\bar{V}(k)), N(k) = g_j^*(k)(N(k)) \mid \bar{A}(k-1) = d_{x_j}^*(\bar{V}(k-1)), \bar{N}(k-1) = g_j^*(\bar{N}(k-1)) \Big)$$

is a stabilizing factor denoting the estimated probability of a patient following both the dynamic treatment-censoring and static monitoring regimen $(d_{x_j}^*, g_j^*)$ at time $k$ given she followed both the treatment-censoring and monitoring interventions $(d_{x_j}, g_j^*)$ through time $k-1$. We note that, unlike the weight function (9), the function $h(t)$ does not involve the terms $P_n(N(k)|\bar{L}(k), \bar{A}(k), \bar{N}(k-1))$ for time points $k$ when $n_j(k) = 0$ in the denominator and the resulting weights $h_i(t)$ can thus be expected to have a more compact distribution and be less prone to take on large values.

The IPW estimators of the counterfactual hazards above are then mapped into an estimator $\psi_n^{(d_{x_1}^*, g_1^*), (d_{x_2}^*, g_2^*)}$ of the causal estimands (10) as follows:

$$\prod_{t=0}^{t_0} \left(1 - P_n(Y_{d_{x_2}^*, g_2^*}(t+1) = 1 | Y_{d_{x_2}^*, g_2^*}(t) = 0)\right) - \prod_{t=0}^{t_0} \left(1 - P_n(Y_{d_{x_1}^*, g_1^*}(t+1) = 1 | Y_{d_{x_1}^*, g_1^*}(t) = 0)\right)$$

and a conservative estimate of its variance is given by a straightforward extension of the results presented in the Appendix of.[20]

We recall that this IPW estimator and its variance estimator are consistent for estimating the causal estimand (10) if the SRA (12) and positivity assumption (13) hold and if the denominators of the IP weight $h(t)$ are consistent estimators of the exposure and monitoring assignment mechanisms, i.e., the true unknown conditional probabilities $P(A(k)|\bar{L}(k), \bar{A}(k-1), \bar{N}(k-1))$ and $P(N(k)|\bar{L}(k), \bar{A}(k), \bar{N}(k-1))$. If the NDE assumption also holds, then this IPW estimator and its variance estimators are also consistent for estimating the causal estimand (6).

### 4.3.1 Implementation

The exact same estimates (i.e., same logistic model fits) of the propensity scores for the exposure and monitoring variables that are used to implement the denominator of the IP weights (9) are also used to implement the denominator of the IP weights (14). Similarly, the estimates of the probabilities $P_n'$ that define the numerators of the IP weights were derived non-parametrically using proportions of observed events. Thus, the only difference between the IPW estimator implementation here and in Section 3.3 is how estimates of the numerators and denominators of the weights are assembled to define the weights $h_i(t)$ assigned to each outcome over time (i.e., the difference between formulas (9) and (14)). In particular, the set of outcomes that contribute to each IPW estimate (4) is different because it is defined by the set of patients $i$ for a given $t$, $x \in \mathcal{X}$ and $\bar{n} \in \mathcal{N}$ such that $h_i(t) \neq 0$. All stabilized weights were also truncated at 40 as in prior analyses.

## 4.4 Results

Figure 9 shows the numbers of patients following each of 16 joint treatment-censoring-monitoring interventions $(d_x^*, g^*)$ for the first 8 periods and can be compared to their analogs $(d_x, \bar{n})$ in Figure 4. These results confirm that the counts of patients following any of the joint interventions $(d_x^*, g^*)$ is larger than their analogs $(d_x, \bar{n})$, except, as expected, when these two interventions are defined based on $\bar{n}_0$. Indeed, the set of patients following each of the two interventions is then identical at each time point because $g_0^* = \bar{n}_0$ which, in turn, implies $d_x^* = d_x$ (because $V^*(t) = V(t)$ when $g_0^* = \bar{n}_0$). At time 0, Figure 9 indicates that, 1) the number of patients following any given intervention $(d_x^*, g^*)$ defined by $\bar{n}_j$ with $j \neq 0$ is identical whichever the definition of $g^*$ (i.e., the value of $j$) and, 2) for each $x \in \mathcal{X}$, this number is equal to the number of patients following the intervention $d_x$ as shown by comparison with the top left panel of Figure 2. These results are expected because $V^*(0) = V(0)$ for all $\bar{n}$ and because the interventions $g^*$ defined by $\bar{n}_j$ for $j = 1, 2, 3$ do not enforce an intervention on monitoring at time 0 and, as a result, the intervention $d_x$ is equivalent to the three interventions $(d_x^*, g^*)$ at time 0. Figure 9 also indicates that, at time 1, the number of patients following any given intervention $(d_x^*, g^*)$ defined by $\bar{n}_j$ with $j = 2, 3$ is identical whichever the definition of $g^*$ (i.e., the value of $j$) and, for each value of $x \in \mathcal{X}$, this number is very close but slightly different from the number of patients following the intervention $d_x$ as shown by comparison with the top left panel of Figure 2. This result reflects the fact that both the interventions $g^*$ defined by $\bar{n}_j$ with $j = 1, 2$ do not enforce an intervention on monitoring at $t = 0, 1$ and thus the difference between intervention $(d_x^*, g^*)$ and $d_x$ at time 1 comes down to the difference between applying the same decision rule $d_x$ when any A1c collected at time 1 is ignored versus used to determine treatment at time 1. The same results and explanation apply to time 2 with $j = 3$ as shown by examining Figure 9 and comparing it to the top left panel of Figure 2. Figure 9 indicates that, starting at time 2 and for each $x \in \mathcal{X}$, the number of patients following each joint intervention $(d_x^*, g^*)$ defined by $\bar{n}_j$ almost always increases with the decrease in the frequency of A1c monitoring (i.e., with the increase in $j$). This result differs from that observed for interventions $(d_x, \bar{n})$ in Figure 4 and demonstrates that the expected increase in the number of patients who follow the monitoring intervention $g^*$ defined by $\bar{n}_j$ as $j$ increases (because it enforces interventions on a smaller subset of monitoring events) is not offset in this study by a decrease in the number of patients who follow the intervention $d_x^*$ over $d_x$.

We contrast the counterfactual survival curves denoted by the 16 joint treatment-censoring-monitoring interventions $(d_x^*, g^*)$, first by comparing the four dynamic treatment-censoring interventions when they are all combined with the same monitoring intervention (Figures 10 and 11), and second by comparing the four monitoring interventions when they are all combined with the same dynamic treatment-censoring intervention (Figures 12 and 13). We note that the equivalence between intervention $(d_x, \bar{n}_0)$ and $(d_x^*, g_0^*)$ for each $x \in \mathcal{X}$ described above explains that the plots on the top left panel of Figures 5 (resp. 6) and 10 (resp. 11) are identical. This equivalence also explains why, on all panels of Figures 7 and 12, the red survival curves corresponding with the monitoring intervention $\bar{n}_0$ and $g_0^*$, respectively, are identical. The increase in the number of patients following the joint interventions $(d_x^*, g^*)$ compared to $(d_x, \bar{n})$ described earlier explains the large decrease in the standard errors of the IPW risk difference estimators (i.e., the width of confidence intervals) displayed in Figures 11 and 13 compared to those displayed in Figures 6 and 8, respectively. The large increase in available information when $g^*$ is defined based on $\bar{n}_j \neq \bar{n}_0$ leads to survival curves estimates on Figure 11 that are much smoother than that displayed on Figure 6. There is now a clear separation and consistent ordering of the survival curve estimates over time for all three interventions $g^*$ defined by $\bar{n}_j$ with $j \neq 0$ and results are now very similar to the estimates obtained without monitoring intervention (bottom panel of Figure 2). Whichever the definition of the intervention $d_x^*$, results from the joint treatment-censoring-monitoring interventions $(d_x^*, g^*)$ indicate that the risk of onset or progression of albumbinuria decreases with the decrease of the A1c threshold at which an intensified treatment is initiated. Unlike evidence from Figure 6, the precision in the risk difference estimates displayed in Figure 11 now provides strong statistical evidence of a protective effect of treatment intensification at a lower A1c threshold when $g^*$ is defined based on $\bar{n}_j$ with $j = 1, 2, 3$. When contrasting joint interventions defined by different monitoring interventions but the same dynamic treatment-censoring intervention on Figure 12, results suggest a decreased risk of onset or progression of albuminuria for any given treatment-censoring intervention $d_x^*$ with $x \in \mathcal{X}$ when the monitoring intervention $g^*$ is defined by $\bar{n}_j$ with $j$ increasing. This result can be approximated by stating that more frequent A1c monitoring results in improved outcome. We note that the estimates of this suggested beneficial effect tend however to be smaller than those displayed on Figure 7 for intervention $(d_X, \bar{n})$. Despite the tighter confidence intervals for the risk difference on Figure 13 compared to those on Figure 8, the general decrease in the point estimates of the risk differences lead to similarly weak statistical evidence of a protective effect of more frequent A1c monitoring.

Table 5 displays the distribution of the stabilized and untruncated IP weights for person-time-regime outcomes during the first 8 quarters of follow-up with nonzero weight values. We note that the distribution of the IP weights is shifted left compared to that for interventions $(d_x, \bar{n})$ displayed in Table 3 but that it remains slightly shifted right compared to that for regimes without monitoring interventions displayed in Table 2. A cursory comparison of the proportions of large weights between Tables 5 and 3 suggests no worsening in near-violations of the positivity assumption. Indeed, the number of weights greater than or equal to 40 and 150 less than doubles even though the number of regime-person-time observations more than doubles when evaluating joint interventions $(d_x^*, g^*)$ instead of $(d_x, \bar{n})$. When excluding observations associated with interventions $(d_x^*, g^*)$ defined by $\bar{n}_j$ with $j = 0$ (i.e., interventions that are then equivalent to $(d_x, \bar{n}_0)$), Table 6 shows a relatively large increase in the proportions of weights greater than or equal to 40 and 150 compared to Table 4. This increase in the number of large weights despite the increase of the number of patients following the interventions $(d_x^*, g^*)$ over $(d_x, \bar{n})$ confirms the theoretical concerns that we detailed earlier over increased near-violations of the positivity assumption requirements for the exposure process in (13) compared to (8). These secondary data analyses demonstrate that efficiency gains that result from greater numbers of patients following interventions $(d_x^*, g^*)$ compared to interventions $(d_x, \bar{n})$ could be offset by increased finite-sample bias resulting from near-violations of the positivity assumption.

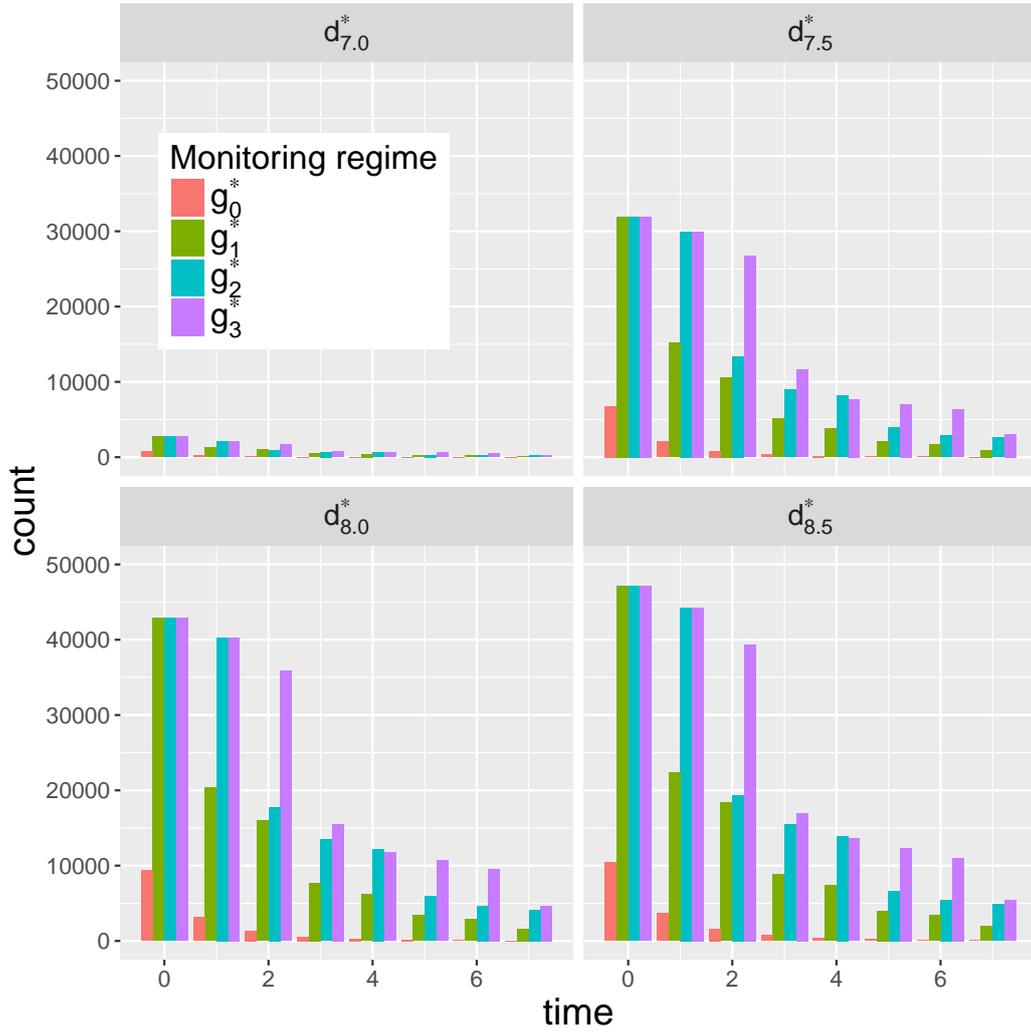

Figure 9: Counts of patients following each of the 16 joint treatment-censoring and monitoring interventions $(d_x^*, g_j^*)$ over time. Counts are organized by dynamic treatment-censoring interventions. The static monitoring interventions $g_j^*$ on the subset of the monitoring process is defined by the monitoring regime $\bar{n}_j$ that requires that two consecutive A1c tests always be separated by $j$ period(s).

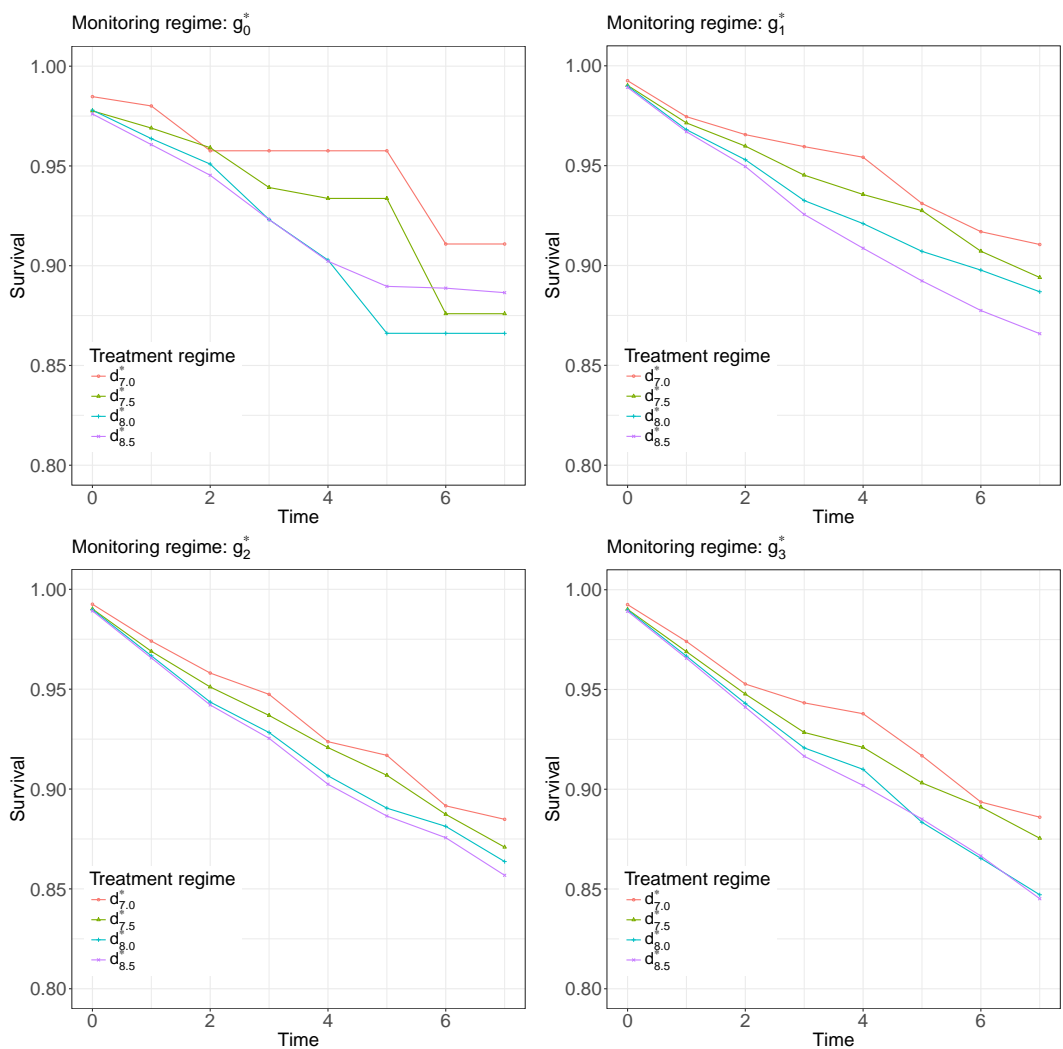

Figure 10: IPW estimates of the counterfactual survival curves under the 16 joint dynamic treatment-censoring and monitoring interventions $(d_x^*, g_j^*)$. Each plot contrasts the four counterfactual survival curves defined by four distinct dynamic treatment-censoring interventions and a single static monitoring intervention. The static monitoring intervention $g_j^*$ on the subset of the monitoring process is defined by the monitoring regime $\bar{n}_j$ that requires that two consecutive A1c tests always be separated by $j$ period(s).

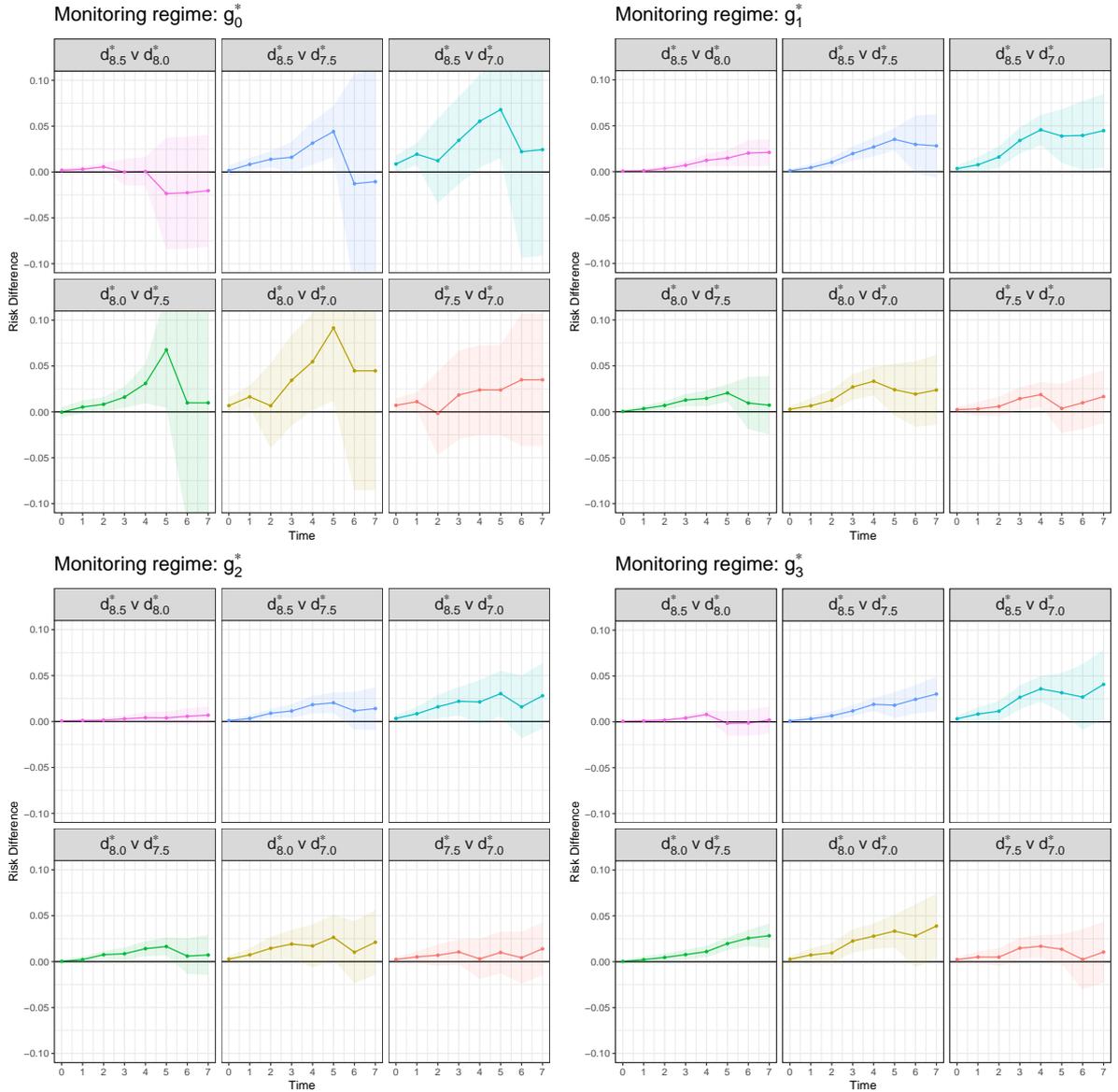

Figure 11: IPW estimates of counterfactual risk differences between dynamic treatment-censoring regimes $d_x^*$ combined with a given static monitoring intervention $g_j^*$. The shaded areas represent the 95% confidence intervals. The static monitoring intervention $g_j^*$ on the subset of the monitoring process is defined by the monitoring regime $\bar{n}_j$ that requires that two consecutive A1c tests always be separated by $j$ period(s).

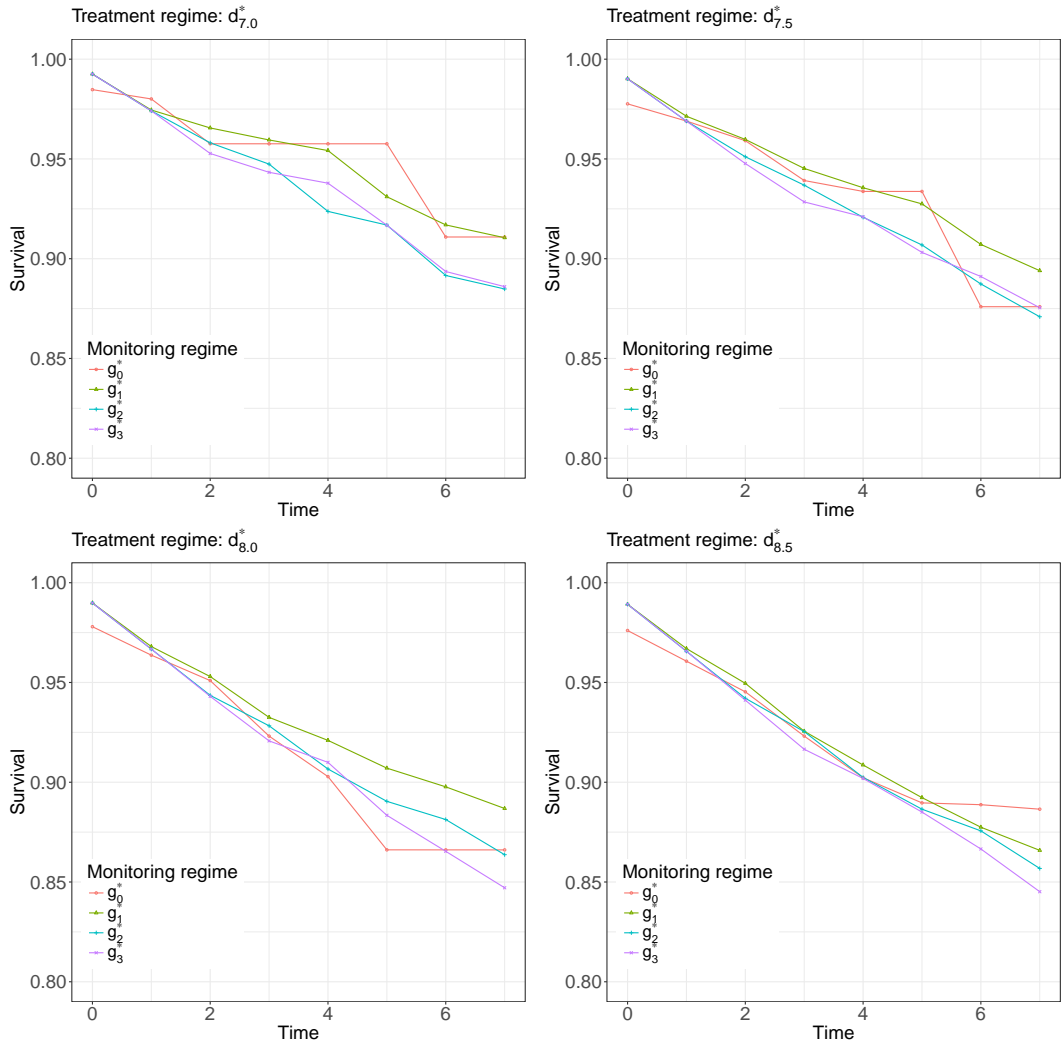

Figure 12: IPW estimates of the counterfactual survival curves under the 16 joint dynamic treatment-censoring and monitoring interventions $(d_x^*, g_j^*)$. Each plot contrasts the four counterfactual survival curves defined by four distinct static monitoring interventions and a single dynamic treatment-censoring intervention. The static monitoring intervention $g_j^*$ on the subset of the monitoring process is defined by the monitoring regime $\bar{n}_j$ that requires that two consecutive A1c tests always be separated by $j$ period(s).

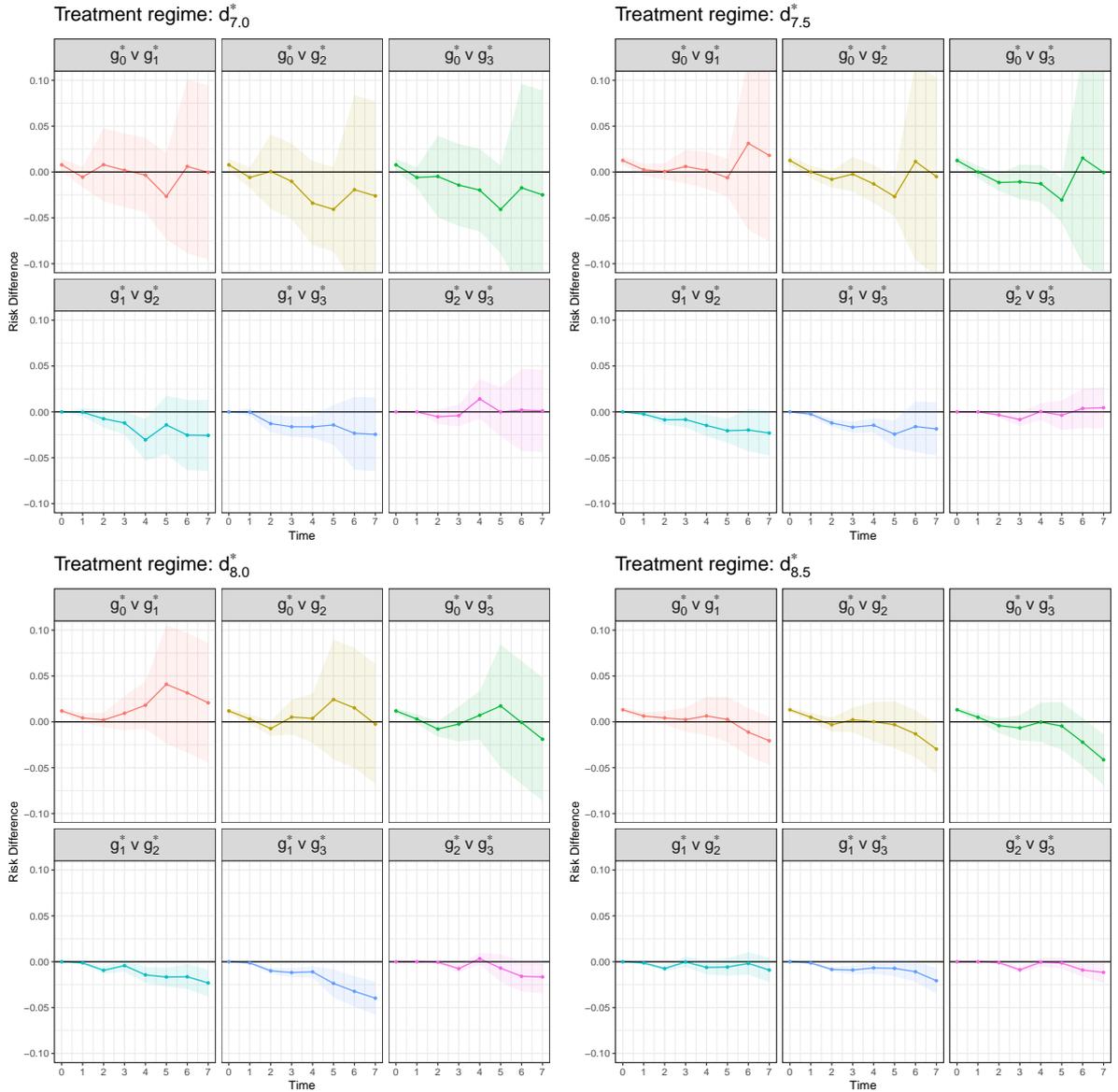

Figure 13: IPW estimates of counterfactual risk differences between static monitoring regimes $g_j^*$ combined with a given dynamic treatment-censoring intervention $d_x^*$. The shaded areas represent the 95% confidence intervals. The static monitoring intervention $g_j^*$ on the subset of the monitoring process is defined by the monitoring regime $\bar{n}_j$ that requires that two consecutive A1c tests always be separated by $j$ period(s).

Table 5: Distribution of stabilized inverse probability weights for evaluating causal estimand (10). Rule-person-time observations with a weight value of 0 are excluded.

| IPW | Frequency | % | Cumulative Frequency | Cumulative % |
|---|---|---|---|---|
| $<0$ | 0 | 0.00 | 0 | 0.00 |
| $[0, 0.5[$ | 105209 | 8.49 | 105209 | 8.49 |
| $[0.5, 1[$ | 952594 | 76.90 | 1057803 | 85.40 |
| $[1, 10[$ | 174362 | 14.08 | 1232165 | 99.47 |
| $[10, 20[$ | 5400 | 0.44 | 1237565 | 99.91 |
| $[20, 30[$ | 752 | 0.06 | 1238317 | 99.97 |
| $[30, 40[$ | 192 | 0.02 | 1238509 | 99.99 |
| $[40, 50[$ | 59 | 0.00 | 1238568 | 99.99 |
| $[50, 100[$ | 93 | 0.01 | 1238661 | 100.00 |
| $[100, 150[$ | 13 | 0.00 | 1238674 | 100.00 |
| $\geq 150$ | 20 | 0.00 | 1238694 | 100.00 |

Table 6: Distribution of the subset of stabilized inverse probability weights for evaluating causal estimand (10) when $\bar{n}_1$ and $\bar{n}_2$ do not enforce continuous A1c monitoring. Rule-person-time observations with a weight value of 0 are excluded.

| IPW | Frequency | % | Cumulative Frequency | Cumulative % |
|---|---|---|---|---|
| $<0$ | 0 | 0.00 | 0 | 0.00 |
| $[0, 0.5[$ | 100184 | 8.38 | 100184 | 8.38 |
| $[0.5, 1[$ | 922681 | 77.21 | 1022865 | 85.60 |
| $[1, 10[$ | 166157 | 13.90 | 1189022 | 99.50 |
| $[10, 20[$ | 5040 | 0.42 | 1194062 | 99.93 |
| $[20, 30[$ | 652 | 0.05 | 1194714 | 99.98 |
| $[30, 40[$ | 149 | 0.01 | 1194863 | 99.99 |
| $[40, 50[$ | 36 | 0.00 | 1194899 | 100.00 |
| $[50, 100[$ | 40 | 0.00 | 1194939 | 100.00 |
| $[100, 150[$ | 6 | 0.00 | 1194945 | 100.00 |
| $\geq 150$ | 7 | 0.00 | 1194952 | 100.00 |

# 5 Discussion

With this case study of the evaluation of adaptive treatment strategies with EHR data, we aimed to provide detailed practical guidance on how to 1) approach the challenges posed by the non-systematic monitoring of time-varying covariates for confounding adjustment and, 2) exploit such monitoring variability to evaluate its health impact or to improve the generalizibility of CER findings. While the approaches developed were illustrated with EHR data, they are applicable to any observational study with non-systematic covariate monitoring.

First, using DAG reasoning, we demonstrated that covariate information that was not systematically collected during follow-up does not necessarily amount to partially missing data, i.e., data that are needed to adequately account for confounding bias and that, if ignored, would lead to a violation of the SRA. Instead, we argued that, under an assumption that is often reasonable in practice, the effects of treatment interventions (dynamic or static) can be identified simply by including an indicator of monitoring events for the partially observed covariate in the adjustment set. For instance, a laboratory measurement that was not ordered by a clinician and that was thus unknown to her cannot influence her treatment decision (e.g., prescribe a new drug). Consequently, imputation of laboratory measurements that were not ordered is unnecessary and adjustment for the time-varying indicator of laboratory monitoring is sufficient for proper confounding adjustment. Our analytic example demonstrated that the use of indicators for covariate monitoring events to estimate IP weights does not impede proper confounding adjustment and that it can even lead to more precise IPW effect estimates. The rationale above questions the relevance of common practice that treats partially unobserved covariate information as missing and attempts to impute it. We argued that even if the imputation of this information is successful, this common approach may not address confounding by the intensity of monitoring except if it also involves adjustment for indicators of monitoring events. Thus, we propose that this imputation approach be instead reserved for covariate measurements that were known to the decision maker when choosing treatment but that were not captured in the data assembled for research purposes. With this perspective, the common practice of defining time-varying covariates by LOVCF should not be construed as an approach for missing data imputation but merely a pragmatic approach to define time-varying covariates that capture the last information known to the decision maker when choosing a treatment. When used for this purpose, LOVCF should thus not be subject to the (valid) criticisms that this approach has been the target of in the context of missing data imputation such as for the imputation of missing outcomes due to, for instance, interval censoring (43, MD-2 PCORI standard). We hope that our results will discourage blanket recommendations against the use of LOVCF in CER and promote more nuanced guidelines that acknowledge its utility for defining time-varying covariates used for confounding adjustment in CER with longitudinal observational data. More importantly, we hope that these results will also encourage routine adjustment for indicators of monitoring events in IPW analyses or their analogs based on alternate estimators (e.g., G-computation, augmented-IPW, or targeted minimum loss based estimators).

Second, because the monitoring of the time-varying covariate(s) that are used to adapt treatment decisions plays a central role in the evaluation of dynamic regimes, we described and compared the implementation of two IPW estimators for evaluating the joint effects of dynamic treatment-censoring and static monitoring interventions. To our knowledge, this report provides the first detailed account of the practical trade-offs between these two IPW estimation approaches. Motivations for their applications include the joint optimization of treatment and monitoring decisions, improved generalizability of study findings, or the evaluation of the health impact of various monitoring strategies when combined with a given dynamic treatment intervention. We note that the performance of these estimators hinges on a stronger SRA whose upholding has been questioned when determinants of monitoring events are not routinely collected on all subjects such as in EHR-based studies. With our example, we illustrated the expected poor performance of standard IPW estimation of these joint effects due to a large decrease in data support to evaluate them which, in turn, can also increase concerns over finite-sample bias from near-violations of the positivity assumption for the monitoring process. To alleviate the expected practical limitation of the standard IPW estimator, we demonstrated how an alternate approach that hinges on an NDE assumption can result in much improved estimation efficiency due to increased data support but at the cost of a potential increase in finite-sample bias due to structural near-violations of the positivity assumption for the treatment process. Evidence from the implementation of these two IPW estimators in the TI study suggests a beneficial effect of more frequent A1c monitoring when combined with a dynamic intervention that requires that an intensified treatment be initiated when A1c reaches or drifts above a threshold >7%.

Results from the evaluation of the joint effects of the four static monitoring interventions and the dynamic treatment-censoring intervention $d_{7.0}$ suggest a possible violation of the NDE assumption in the TI study. Indeed, the dynamic rule $d_{7.0}$ is equivalent to the static treatment intervention $\bar{a} = (1, \ldots, 1)$ because, by study design, all patients enter the cohort with a baseline A1c measurement $\geq 7\%$. Under the NDE assumption (11), we would thus expect little to no differentiation between the survival curves in the top left panel of Figure

7. Instead, these curves indicate a consistent increase in the estimated risk of failure with the decrease in the frequency of A1c monitoring although these apparent differences generally do not reach statistical significance as shown on the top left panel of Figure 8. We note that these estimates could also result from unmeasured confounding of the effect of A1c monitoring on the outcome, i.e., the NDE assumption could actually hold but biased survival estimates make it appear as violated. Even if the NDE assumption were violated, we emphasize that inferences from the alternate NDE-based IPW analyses in this paper can remain causally interpretable although results become more difficult to convey and use to inform care.

As noted in (9, p. 4703), we might expect a violation of the NDE assumption in intention-to-treat (ITT) analyses where the dynamic treatment interventions stop after initiation of an intensified therapy but the monitoring interventions continue past treatment intensification. For instance in the TI study, the NDE assumption would be violated in such an analysis if A1c monitoring motivates patient adherence, i.e., the refill of prescriptions for the intensified therapy later after its initiation. Our analyses were based on the per-protocol principle, i.e., the dynamic treatment intervention continues after initiation of an intensified treatment by requiring that patients remain on the intensified therapy thereafter. We note that an alternate approach to mitigate concerns over NDE violations in ITT analyses could consist in replacing the current static monitoring interventions that continue through outcome collection with equivalent interventions that stop monitoring requirements after the intensified treatment is initiated. Such an approach would however require extending the identifiability results in (15).

Finally, we note that an alternate analytic strategy that would not rely on the NDE assumption to address the practical limitations of standard IPW estimators of joint treatment-monitoring effects could consist in evaluating stochastic monitoring interventions instead of static ones. These alternate interventions can lead to the definition of joint effects that are particularly relevant for patient-centered outcomes research because they can better represent real-world adherence to rigid monitoring schedules such as the ones studied in this report. The weakening of the positivity assumption required to identify these effects is expected[44] to improve the practical performance of their standard IPW estimators compared to that of the standard IPW estimators of effects defined by static monitoring interventions.